\documentclass[11pt]{article}

\usepackage{jheppub}

\usepackage{amsmath, amssymb}
\usepackage[english]{babel}
\usepackage[utf8]{inputenc}
\usepackage{enumitem}
\usepackage{latexsym}
\usepackage{graphicx}
\usepackage{slashed}
\usepackage{xcolor}
\usepackage{color}
\usepackage{float}
\usepackage{bm}
\usepackage{braket}
\usepackage{dsfont}
\usepackage{cancel}
\usepackage{mathrsfs}
\usepackage{tikz}

\usepackage{caption}
\usepackage{subcaption}

\usepackage{hhline}

\title{\bf TransPlanckian Censorship on Dark Dimension Inflation}

\author[a,b,c]{Luis A. Anchordoqui,}

\affiliation[a]{Department of Physics and Astronomy,  Lehman College, City University of
  New York, NY 10468, USA
}

\affiliation[b]{Department of Physics,
 Graduate Center,  City University of
  New York,  NY 10016, USA
}

\affiliation[c]{Department of Astrophysics,
 American Museum of Natural History, NY
 10024, USA
}

\author[d,e]{Ignatios Antoniadis,}

\affiliation[d]{School of Natural Sciences, Institute for Advanced Study, Princeton, NJ 08540, USA}

\affiliation[e]{Laboratoire de Physique Th\'eorique et Hautes \'Energies
  - LPTHE 
Sorbonne Universit\'e, CNRS, 4 Place Jussieu, 75005 Paris, France
(present address)
}

\author[f]{and Jules Cunat,}

\affiliation[f]{High Energy Physics Research Unit, Faculty of Science, Chulalongkorn University, Bangkok 1030, Thailand}

\abstract{It was proposed that extra dimensions can acquire large size by higher dimensional inflation connecting two large hierarchies in particle physics and cosmology, namely the weakness of the actual gravitational force to the largeness of the observable Universe, in terms of one fundamental scale. This proposal is consistent with the observed approximate scale invariant power spectrum of primordial density perturbations only for one or two extra dimensions of around the micron size. While a cosmological history connecting the period of higher dimensional inflation to the beginning of the standard cosmology has recently been studied, we investigate here the TransPlanckian Censorship Conjecture in that context and show that it drastically constrains the parameter space of the model.
}

\begin{document}
\maketitle

\section{Introduction}

It is well-known that cosmic inflation provides an economic and elegant solution to the horizon and
flatness problems of the standard hot big bang model by postulating a
period of rapid, exponential expansion in the early
universe~\cite{Guth:1980zm,Starobinsky:1980te,Linde:1983gd}. Nevertheless,
it is also a well-established fact that this framework faces a perplexing
puzzle known as the
{\it trans-Planckian problem}~\cite{Martin:2000xs}. During inflation,
quantum fluctuations are exponentially stretched. If
inflation lasts long enough, the wavelengths of the primordial
perturbations we could observe in the cosmic microwave background (CMB)
would have originated at length scales shorter than the Planck length
($\ell_p \sim 1.6 \times 10^{-35}~{\rm m}$), where our current
understanding of Einstein's gravity breaks down and quantum gravity
takes over. Obviously, to even make this claim, we  have to assume that we can reliably calculate how quantum
perturbations behave even when they originate beyond the limits of
effective quantum field theory.

The trans-Planckian censorship conjecture (TCC) attempts to resolve the trans-Planckian problem by postulating that in
any consistent theory of quantum
gravity, fluctuations with sub-Planckian length scales are strictly
forbidden from exiting the Hubble horizon~\cite{Bedroya:2019snp}. The
TCC then implies that all information about trans-Planckian scales
remains hidden from the {\it classical domain}, the domain where
fluctuations grow and {\it can} classicalize, i.e. the super-Hubble
region. More concretely, in a uniformly expanding universe with scale factor $a(t)$ and
Hubble parameter $H(t) = \dot a/a$ the TCC dictates that
\begin{equation}
  \frac{a(t_f)}{a(t_i)} \equiv \frac{a_f}{a_i} <
  \frac{M_p}{H_f} 
\label{censrule}
\end{equation}
for any initial and final times $t_i < t_f$, where $M_p$ is the reduced
Planck mass. In plain English, the TCC states that the universe is {\it censored}
from the possibility of observing the chaotic, quantum-scale physics of gravity;
no sub-Planckian quantum fluctuation can be stretched to a
cosmological (macroscopic) size.

The TCC falls under the umbrella of the Swampland Program~\cite{Vafa:2005ui}, an effort
to identify which low-energy effective field theories are consistent
with nonperturbative quantum gravity considerations, versus the set of
consistent-looking effective field theories that do not admit a UV
completion in quantum gravity (and are said to live in the {\it
  swampland})~\cite{Palti:2019pca,vanBeest:2021lhn,Agmon:2022thq}.  In particular, the censorship rule (\ref{censrule})
sets a strict upper bound on the maximum duration of
inflation. Because the universe expands so rapidly, the total number
of e-folds during inflation  $N = \log (a_f/a_i)$ is
constrained to be
\begin{equation}
 N = \int_{t_i}^{t_f} H \ dt < \log \frac{M_p}{H_f} \, ,
\label{TCCefolds}
\end{equation}  
where we have used $\int H(t) \ dt = \log(a_f/a_i)$
and where $H_f$ is the Hubble
rate at the end of inflation~\cite{Bedroya:2019snp}. 

Applying the TCC imposes severe constraints on cosmic
inflation~\cite{Bedroya:2019tba,Brandenberger:2021pzy,Bedroya:2024zta,Bedroya:2025ris,Anchordoqui:2026hys}. For
example, assuming immediate reheating after inflation, (\ref{censrule})
leads to an upper bound on the scale of inflation $H_I$
\begin{equation}
  H_f \equiv H_I \lesssim 10^{-20} M_p \,,
  \label{Hinfbound}
\end{equation}
which forces the tensor-to-scalar ratio $r$ (a measure of primordial
gravitational waves) to be negligibly small~\cite{Bedroya:2019tba}
\begin{equation}
  r = \frac{2}{\pi^2 \ A_s}\left(\frac{H_I}{M_p}\right)^2 <
    6.8 \times 10^{-33} \, ,
\label{r}
\end{equation}
where $A_s \approx 2.1 \times 10^{-9}$ is the amplitude of the primordial power
spectrum of scalar density
fluctuations~\cite{Planck:2018vyg}. Relaxing the assumption of
instantaneous reheating post-inflation eases the constraint (\ref{r})
to $r \lesssim 10^{-10}$~\cite{Mizuno:2019bxy}; however, this still
makes it challenging, if not impossible, to detect primordial B-mode
polarization in the CMB. Moreover, during inflation,
\begin{equation}
H_I^2 \sim \frac{\Lambda_I}{3 M_p^2},
\label{LambdaFriedmann}
\end{equation}
yielding via 
(\ref{Hinfbound}) an upper bound for
the energy scale of inflation~\cite{Brandenberger:2021pzy}
\begin{equation}
\eta = \Lambda_I^{1/4}= \left(3H_I^{2}M_{p}^{2} \right)^{1/4} \lesssim 10^9~{\rm GeV}
\, .
\end{equation}
Thus, to satisfy the TCC bound, the energy scale of inflation must be
dramatically lower than the traditional Grand Unified Theory (GUT)
scale (of $10^{16}~{\rm GeV}$). In addition, models that obey the TCC require significant fine-tuning of the initial conditions before inflation begins, such as demanding that the inflaton field's initial velocity is almost perfectly zero~\cite{Bedroya:2019tba}.

In standard 4D cosmology, the energy scale of inflation is typically
near the GUT scale. In large extra dimension scenarios, $M_p$ is an
emergent phenomenon linked to the 
higher dimensional volume ${\cal V}$~\cite{Arkani-Hamed:1998jmv,Antoniadis:1998ig}. Because
the fundamental gravity scale $M_*$ (or species scale where gravity becomes
strong) could be extremely low for large ${\cal V}$
\begin{equation}
M_*  \sim  M_p^{2/(2+d)}/{\cal V}^{1/(2+d)} 
\end{equation}
 the natural scale for the inflationary potential is drastically
lowered, allowing inflation to occur at $\eta \sim 10^9~{\rm GeV}$ or
even below~\cite{Dvali:1998pa}. Throughout this paper we align with the
dark dimension scenario~\cite{Montero:2022prj} and consider extra dimensions compactified on
line intervals in which ${\cal V} = (\pi R_\perp)^d$, with $d$ the
number of large extra dimensions and $R_\perp $ of micron scale. The
only two possibilities consistent with collider experiments and
astrophysical observations are $d=1$ where $M_* \sim 10^9~{\rm GeV}$,
and $d = 2$ where $M_* \sim
10~{\rm TeV}$~\cite{Anchordoqui:2025nmb}. Cosmological bounds, however, suggest borderline
experimental feasibility for $d=2$~\cite{Hannestad:2001nq}.  The
anti-de Sitter distance conjecture extended to de Sitter space~\cite{Lust:2019zwm} links the micron-scale extra
dimensions to the cosmological hierarchy problem, $\Lambda/M_p^4 \sim
10^{-120}$. Thus, the dark dimension provides a theoretical framework
where the tiny value of the cosmological constant $\Lambda$ sets the size of the compactified dimensions~\cite{Montero:2022prj}.

Now, in de Sitter (dS) spacetime with a radius of $1/H_I$, massive
spin-2 particles are subject to the Higuchi bound, which dictates a
strict minimum mass of
$m^2 \geq 2H_I^2$~\cite{Higuchi:1986py}. Dropping below this lower
limit introduces negative-norm helicity modes, which directly violates unitarity. Because the characteristic size of the extra
dimensions corresponds to the inverse mass of the lightest
Kaluza-Klein (KK) graviton excitation ($m_{\rm KK} \sim 1/R_\perp$), the Higuchi bound effectively dictates
that the extra dimensions cannot exceed the dS horizon $1/H_I$ by more
than an $\mathcal{O}(1)$ factor. This bound then forces a very
small $H_I \leq {\rm eV}$ for inflation at fixed large (micron scale)
extra dimension~\cite{Anchordoqui:2022svl}.\footnote{This bound was already imposed in~\cite{Dvali:1998pa} using a different argument.}

These challenges can be resolved by assuming the universe underwent an
inflationary period where the radius of the dark dimension expanded
exponentially, stretching from the species length ($R_0 \sim
M_*^{-1}$) up to the micron scale. Higher dimensional slow-roll
inflation naturally yields an approximately scale invariant power
spectrum for primordial density perturbations, aligning with CMB
observations~\cite{Anchordoqui:2023etp}. This alignment occurs because the two-point
function of a massless, minimally coupled scalar field in dS space
scales logarithmically past the cosmological horizon, independent of
spacetime dimensionality~\cite{Ratra:1984yq}. However, in compactified models, this
scale invariance typically breaks down at distances larger than the
compactification length, potentially creating tensions with
large-scale CMB data. Nonetheless, in a  companion
paper~\cite{Anchordoqui:2026los} we have demonstrated that for $d =
1$, the predictions from five-dimensional (5D) inflation remain consistent with
 Planck data~\cite{Planck:2018nkj,Petretti:2024mjy} and fall within
 the cosmic variance error band. In this paper we investigate the TCC
 bounds on higher dimensional inflation, focusing on both $d=1$ and $d=2$.

 The layout of the paper is as follows. In Sec.~\ref{sec:TCC}, we review how the implications of the TCC can be relaxed if reheating is delayed and does not occur immediately after the end of inflation. In Sec.~\ref{sec:4Dinf}, we analyze the constraints on inflation in the presence of a stabilized large extra dimension. In particular, we show that the Higuchi bound imposes a stronger constraint than the TCC, leading to an extremely strong suppression of primordial gravitational waves, as it requires $r\lesssim10^{-50}$. We therefore turn to the case of five-dimensional inflation in Sec.~\ref{setup}, allowing the fifth dimension to dynamically grow during inflation and thereby evade the Higuchi bound. In Sec.~\ref{sec:5}, we discuss the formulation of the TCC in the context of five-dimensional inflation and analyze how it constrains the parameter space of the model. As the results of this analysis show that both the Hubble radius and the size of the extra dimension must be much larger than the fundamental length $M_*^{-1}$ at the onset of inflation, we discuss the possibility of a pre-inflationary phase responsible for generating these initial conditions in Sec.~\ref{sec:6}. We then generalize the analysis to the case of two extra dimensions in Sec.~\ref{sec:7} and show that this scenario is not phenomenologically viable once the TCC is imposed. Finally, in Sec.~\ref{sec:Mtheory}, we discuss how the dark dimension scenario could be naturally embedded within M-theory, providing a realization of the pre-inflationary phase introduced in Sec.~\ref{sec:6}. We conclude in Sec.~\ref{sec:conclusion}.

\section{Relaxing TCC constraints through delayed post-inflationary
  reheating \label{sec:TCC}}

As anticipated in the Introduction, the TCC limits can be loosened in
 inflationary models that lack an immediate reheating phase. To
 facilitate the subsequent discussion, we briefly review how this concept takes shape.

Implementing this idea is straightforward. While the TCC imposes an
upper bound on the number of inflationary e-folds,
\begin{align}
  e^{N} \leq \frac{M_p}{H_I} \,,
\label{TCCeN1}
\end{align}  
resolving the horizon problem requires the comoving size of the
today's observable universe to be smaller than the comoving Hubble
radius at the start of inflation
\begin{align}\label{horizoncon}
  \frac{D_{\text{LSS}}}{a_t}\lesssim\frac{1}{a_iH_I} \,,
\end{align}
where  $D_{\rm LSS} \simeq 14~{\rm Gpc}$  indicates today's proper distance to the
last scattering surface (LSS), while $a_t$ and $a_i$ denote the scale
factors today and at the onset of inflation,
respectively. Equation~(\ref{horizoncon}) can be recast as
    \begin{align}
D_{\text{LSS}}\lesssim e^{N_r+\Delta N+N}\frac{1}{H_I}
    \end{align}
where $N_r$ is the number of e-folds from reheating to the present day and
$\Delta N$ is the number of e-folds between the end of inflation and
reheating.

By combining the TCC constraint with the requirement to solve the
horizon problem without assuming instant reheating, we obtain 
\begin{align}\label{eq:TCC+horizon4D1} 
H_I^2\leq\frac{M_p}{D_{\text{LSS}}}e^{N_r+\Delta N} \, . 
\end{align}
The bound on the inflationary scale (\ref{Hinfbound}) can be relaxed
by maximizing the expansion during the post-inflationary eras. This optimization
requires two conditions. First, a low reheating temperature minimizes
the expansion from reheating to today. The scaling of $N_r$ with the
reheating temperature has been computed
in~\cite{Kreling,German:2020kdp} and is given by
\begin{align}\label{eq:N_r}
  e^{N_r}\simeq e^{30}\frac{T_r}{\text{GeV}}.
\end{align}
Second, a linear expansion phase ($\ddot{a}=0$) between inflation and
reheating, characterized by an equation of state parameter $w=-1/3$,
maximizes the intermediate growth~\cite{Mizuno:2019bxy}. A motivation and a comprehensive rationale for this phase is deferred to Sec.~\ref{subsec:leU}. Under these two conditions, the expansion during this intermediate phase simplifies to
\begin{align}
    e^{\Delta N}=\left(\frac{H_I}{H_r}\right)^{\frac{2}{3(1+w)}}=\frac{H_I}{H_r}.
\end{align}

Considering a radiation dominant (RD) phase after reheating, one obtains the Hubble parameter at reheating as
\begin{equation}\label{eq:hr}
   H_r=\sqrt{\frac{\pi^2g_*}{90}}\frac{T_r^2}{M_p} \,,
 \end{equation}
where $g_*$ denotes the effective numbers of relativistic degrees of freedom contributing to the energy density of the thermal plasma. The condition \eqref{eq:TCC+horizon4D1} becomes
\begin{align}\label{eq:TCC+hor4D2}
    \frac{H_I}{\text{GeV}}\leq 9.1\times10^{7}\,g_*^{-1/2}\left(\frac{T_r}{\text{GeV}}\right)^{-1}.
\end{align}
In particular, going as low as $T_r=5\,\text{MeV}\gtrsim
T_{\text{BBN}}$, corresponding to $g_*=10.74$ \cite{Husdal:2016haj}
gives $H_I \leq 5.6\times10^9~\text{GeV}$. As noted in the Introduction
such a low inflation scale implies an extremely reduced amplitude of
the primordial gravitational waves, which via (\ref{r}) leads to
\begin{equation}
r\lesssim10^{-10} \, .
\label{r-relaxed}
\end{equation}

In closing, we note that our estimate for the tensor-to-scalar ratio
(\ref{r-relaxed}) differs from the original value reported in~\cite{Mizuno:2019bxy}. This discrepancy arises because we adopted a higher lower bound for $T_r$, aligning our work with lower
limits on $T_r$ imposed by big-bang nucleosynthesis assuming both
radiative and hadronic decays of such massive particles~\cite{Hasegawa:2019jsa, Kawasaki:2000en,Kawasaki:1999na}.

\section{Constraints on 4D inflation in presence of a large extra dimension}\label{sec:4Dinf}

In this section, we investigate inflation in the presence of fixed large
extra dimensions. The fundamental requirement that the inflationary scale must lie below the species scale, $H_I < M_*$, imposes strict upper bounds on the tensor-to-scalar ratio, restricting it to 
\begin{align}
    r\leq8.5\times10^{-12}\left(\frac{R_\perp}{\mu\text{m}}\right)^{-2/3}
\end{align}
for $d=1$, and
\begin{align}
  r\leq2.5\times10^{-21}\left(R_\perp/\mu\text{m}\right)^{-1}
\end{align}  
for $d=2$.

In addition, TCC constraint (\ref{TCCeN1})  can be rewritten as
\begin{align}
    e^{N} \leq \frac{M_*}{H_I}  \, .
\end{align}
Consequently, the TCC enforces a stricter upper limit on the number of
e-folds when a large extra dimension is present, as it suppresses the
species scale. Specifically, the constraint \eqref{eq:TCC+hor4D2}
transforms into: \begin{align}\label{eq:TCC+hor4D3}
                   \frac{H_I}{\text{GeV}}\leq
                   2.7\times10^{-2}\,g_*^{-1/2}\left(\frac{T_r}{\text{GeV}}\right)^{-1}\left(\frac{R_\perp}{\mu\text{m}}\right)^{-1/3}.
\end{align}
Pushing the parameters to the minimum allowable reheat
                 temperature, $T_r=5~\text{MeV}\gtrsim T_{\text{BBN}}$
                 (where $g_{*}=10.74$~\cite{Husdal:2016haj}), yields
                 the following bound on the tensor-to-scalar ratio
\begin{align}
                   r\lesssim4.4\times10^{-29}\left(\frac{R_\perp}{\mu\text{m}}\right)^{-2/3}
                   \, .
\end{align}

Lastly, as noted in the Introduction, 
in the case of 4D inflation in presence of a large extra dimension,
the Higuchi bound must be satisfied~\cite{Higuchi:1986py}. Since KK gravitons behave as
massive spin-2 states from the four-dimensional perspective, their
masses must satisfy  $m_{\text{KK}}^2\geq2H_I^2$ during any dS phase
in order to guarantee positive norm for the helicity-0 polarization
and prevents ghost instabilities. The mass of the lightest KK
graviton scales as $m_{\text{KK}} \sim 1/R_\perp$, consequently,
\begin{align}\label{eq:Higuchi4D}
     \frac{H_I}{\text{GeV}}\leq
  1.4\times10^{-10}\left(\frac{R_\perp}{\mu\text{m}}\right)^{-1} \,,
\end{align}
and via (\ref{r}) we finally arrive at
   \begin{align}\label{eq:r4D}
       r\leq3.3\times10^{-49}\left(\frac{R_\perp}{\mu\text{m}}\right)^{-2}
     \, .
   \end{align}
Prior to summarizing our findings, we highlight two key observations:
{\it (i)} while we investigated constraints from the Higuchi bound for the lightest KK graviton, it is interesting to note that for the heaviest KK
graviton within the effective field theory of mass $\sim M_*$, the
Higuchi bound reduces to the consistency condition $H_I\lesssim
M_*$; {\it (ii)}~in models with large extra dimensions, the {\it
  normalcy temperature} ($T_{N}$) defines the maximum temperature of
the visible universe's radiation bath before overproduction of KK
modes causes the higher dimensional bulk to overclose the
Universe~\cite{Arkani-Hamed:1998sfv}. A rigorous derivation of $T_{N}$ is provided in Sec.~2 of
the accompanying paper~\cite{Anchordoqui:2026los}. We find $T_N \sim 1~{\rm GeV}$ for $d=1$ and
$T_N \sim 5~{\rm MeV}$ for $d=2$. Evidently, $T_N$ sets a strict upper
bound on $T_r$.

\begin{figure}[h]
     \begin{subfigure}[t]{0.5\textwidth}
    \centering\includegraphics[width=0.98\linewidth]{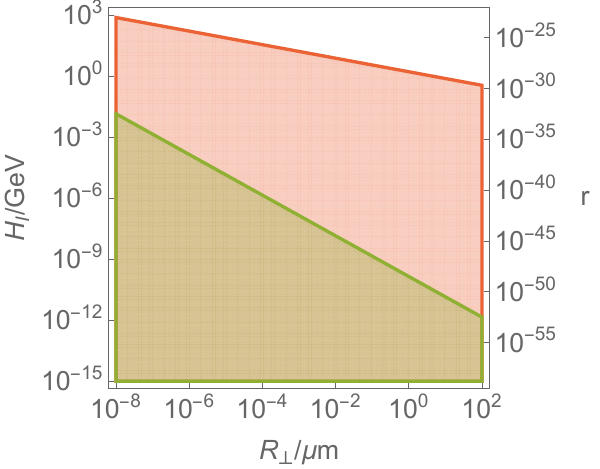}
    \subcaption{$T_r=5\,\text{MeV}\gtrsim T_{\text{BBN}}$}
    \end{subfigure}
    \begin{subfigure}[t]{0.5\textwidth}
    \centering\includegraphics[width=0.98\linewidth]{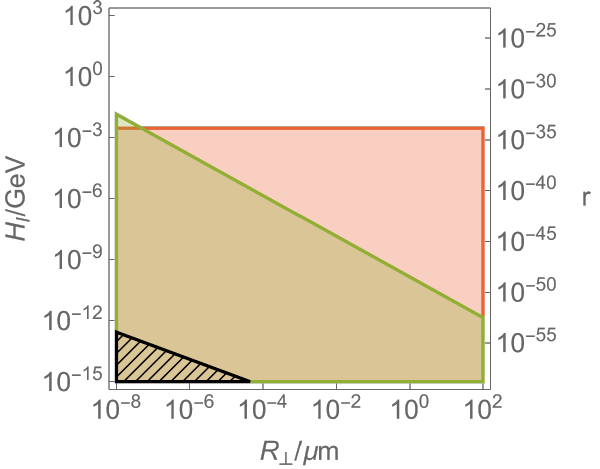}
    \subcaption{$T_r=T_N\simeq(R_\perp/\mu\text{m})^{-1/3}\,\text{GeV}$}
    \end{subfigure}
    \caption{Constraints on 4D inflation in presence of a large extra
      dimension. The green and red regions correspond to the regions
      allowed by the Higuchi bound \eqref{eq:Higuchi4D} and by the TCC
      plus the horizon problem \eqref{eq:TCC+hor4D3},
      respectively. The hatched region is the region excluded by the
      requirement $ H_r \leq H_I$. We use
      $g_*=10.74$ for $T_r=5~\text{MeV}$ and
      $g_*\sim86$ for
      $T_r=T_N$~\cite{Husdal:2016haj}. The corresponding values of the
      tensor-to-scalar ratio
      $r$ are indicated on the right vertical axis.}
    \label{fig:1}
\end{figure}

The constraints on 4D inflation in presence of a large extra dimension
of micron size are encapsulated in Fig.~\ref{fig:1}. For models
featuring a micron-scale dark dimension, the scale of inflation drops
to the eV range or lower, resulting in an exceptionally small
tensor-to-scalar ratio: $r \lesssim 10^{-50}$.

\section{5D inflation}
\label{setup}

In this section, we provide a concise summary of 5D inflation, while an in-depth analysis is detailed in the accompanying paper~\cite{Anchordoqui:2026los}.

\subsection{Setup of 5D inflation and its four-dimensional descriptions}

5D inflation featuring a compact dimension can be modeled using an
approximate dS$_5$ metric. In this framework, both the compact and
non-compact spatial dimensions undergo exponential expansion with
respect to the 5D proper time $\hat{t}$. The corresponding line
element is given by:\begin{align} \label{dS5}
                      ds^2=-d\hat{t}^2+\hat{a}(\hat{t})^2d\vec{x}^2+R(\hat{t})^2dy^2
                      \quad;\quad R(\hat{t})=R_0\,\hat{a}(\hat{t})
                      =R_0\,e^{H_I\hat{t}}\,, \end{align}where the
                    hats explicitly distinguish 5D quantities from
                    their 4D counterparts. The fifth coordinate has a
                    periodicity of $y\sim y+2\pi$ and is compactified
                    on a line interval. This interval is generated via
                    a $Z_{2}$ orbifolding of the circle ($S^1/Z_2$) by
                    identifying the parity $y\to -y$, making
                    $R(\hat{t})$ the physical radius of the
                    circle. The parameters $H_{I}$ and $R_{0}$ define
                    the expansion rate and the initial radius at
                    $\hat{t}=0$ (setting the initial scale factor
                    $\hat{a}_0=1$).\footnote{Note the difference in
                      the normalisation of the scale factor from the
                      usual one $a_t=1$, which is convenient when starting with 5D inflation.} This choice of normalization ($\hat{a}_0=1$) is particularly convenient for analyzing the dynamics of a 5D inflationary phase.

The metric in \eqref{dS5} solves the 5D Einstein field equations in the presence of a positive cosmological constant $\Lambda _{5}^{I}$. The action is expressed as:
\begin{align} \label{S5}     S_5=\int [d^4\vec{x}dy]\left\{\frac{M_*^3}{2}{\cal R}^{(5)}-\Lambda_5^I  \right\}    \quad;\quad 6H_I^2M_*^3 = \Lambda_5^I\,, \end{align}
where $\mathcal{R}^{(5)}$ represents the 5D Ricci scalar and $M_{*}$ denotes the 5D reduced Planck mass. The brackets indicate a coordinate measure that transforms properly as a density under general coordinate transformations. For the effective field theory description to remain valid, both the inflationary Hubble scale $H_{I}$ and the initial compactification scale $1/R_0$ must lie below the fundamental cutoff scale $M_{*}$. Physically, $\Lambda _{5}^{I}$ represents the energy density of a 5D inflaton potential evaluated along a sufficiently flat region to drive 5D slow-roll inflation.

Integrating over the extra dimension $y$ yields the dimensionally
reduced action in the Jordan frame,
\begin{align}     S_J=\int [d^4\vec{x}]\left\{\frac{\pi R
  M_*^3}{2}{\cal R}^{(4)} -\pi R\Lambda_5^I\right\}
\end{align}
where massive KK excitations have been omitted. Although the Jordan
frame lacks an explicit kinetic term for the radion $R$, as the
decomposition of the 5D Ricci scalar $\mathcal{R}^{(5)}$ yields a
total derivative for this component, the radion remains a fully
dynamical degree of freedom. Its dynamics is implicitly encoded
through its non-minimal coupling to gravity via the pre-factor of the
4D Ricci scalar. Performing a Weyl rescaling to transition to the
Einstein frame removes this non-minimal coupling, explicitly
recovering the radion dynamics by generating a canonical kinetic term
(up to field redefinitions). Consequently, 5D inflation can be modeled
via a 4D background framework~\cite{Anchordoqui:2023etp} that solves
the equations of motion of an effective action containing both gravity
and a scalar field (the radion). This effective description is
achieved by integrating over $y$ and rescaling the 4D metric to the
Einstein frame:
\begin{align} \label{S4} 
ds^2 &=\frac{R_\perp}{R(t)}\left\{-dt^2+a(t)^2d\vec{x}^2\right\}+R(t)^2dy^2\nonumber\\  
S_E &=\int [d^4\vec{x}]\left\{\frac{{M}_p^2}{2}\left({\cal R}^{(E)}-\frac{3}{2}\left(\frac{\partial R}{R}\right)^2\right)-\frac{\pi R_\perp^2}{R}\Lambda_5^I\right\}\,, 
\end{align}
where $R_{\perp }$ represents the compactification radius at the end
of
inflation, and the 4D Planck mass is defined by $M_p^2=\pi R_\perp
M_*^3$. In \eqref{S4}, we neglect the massive spin-2 KK
excitations of the 4D graviton because their background values vanish,
and the spin-1 graviphoton zero-mode is eliminated by the $Z_{2}$
parity orbifolding. By defining a canonically normalized radion field
$\sigma$ through the parameterization $R=R_\perp
e^{\sqrt{2/3}{\sigma/M_p}}$, the action takes the standard form:
\begin{align} \label{S4rad}     
S_4=\int [d^4\vec{x}]\left\{\frac{{M}_p^2}{2}{\cal R}^{(4)}-\frac{1}{2}(\partial\sigma)^2 -\pi R_\perp\Lambda_5^Ie^{-\sqrt{\frac{2}{3}}{\frac{\sigma}{M_p}}}\right\} \quad;\quad R=R_\perp e^{\sqrt{\frac{2}{3}}{\frac{\sigma}{M_p}}}\,. 
\end{align}
This setup provides the standard Einstein equations coupled to a scalar field $\sigma$ with an exponential potential $V=\pi R_\perp\Lambda_5^Ie^{-\sqrt{2\over 3}{\sigma\over M_p}}$. Assuming a Friedmann-Lema\^{\i}tre-Robertson-Walker (FLRW) metric $ds_4^2=-dt^2+a^2(t)d\vec{x}^2$ governed by the scale factor $a(t)$, the cosmological equations of motion reduce to:\begin{align} \label{S4EoM} 
3h^2 M_p^2&={1\over 2}\dot{\sigma}^2+V\quad;\quad h={\dot{a}\over a}\\ \nonumber 
2\dot{h} M_p^2&=-\dot{\sigma}^2\,,      \end{align}
where the overdot denotes a derivative with respect to the cosmic time $t$.

The analytical solution to this system can be straightforwardly
derived as:
\begin{align} \label{4Dsolution}
a(t) &= a_e \left( \frac{H_I t}{2} \right)^3, \quad \sigma(t) = \sqrt{6} M_p \ln \left( \frac{H_I t}{2} \right), \quad R(t) = R_\perp \left( \frac{H_I t}{2} \right)^2
\end{align}
where $H_{I}$ is defined in \eqref{S5}, and $a_{e}$ denotes the
scale factor at the end of inflation, which occurs at $t_e = 2 /
H_I$. This solution characterizes a power-law inflationary phase where
the 4D scale factor grows as $a \propto t^3$, while the
extra-dimensional radius expands as $R \propto t^2$. The inflationary
epoch begins when the radius is at its initial value $R_{0}$ at time
$t_{0}$. Consequently, the initial scale factor $a_{0}$ and the
initial radius $R_{0}$ are related by:
\begin{align} \label{a0}
a_0 &= a_e \left( \frac{R_0}{R_\perp} \right)^{3/2}
\end{align}
From this relation, it follows that $N$ e-folds of expansion in the 4D
universe correspond to $\hat{N} = \frac{2}{3}N$ e-folds of radial
expansion driven by the 5D inflation. For convenience in subsequent
analyses, we introduce a dimensionless parameter $\epsilon \equiv
\frac{1}{R_0 M_*} \lesssim 1$. In terms of $\epsilon$, the total
number of inflationary e-folds can be expressed as:
\begin{align} \label{eq:4Defolds}
N &\equiv \ln \left( \frac{a_e}{a_0} \right) = \frac{3}{2} \ln \left( \frac{R_\perp}{R_0} \right) = \frac{3}{2} \hat{N} \\ \label{eq:4Defolds2}
&= \ln(R_\perp M_p) + \frac{1}{2} \ln \left( \frac{\epsilon^3}{\pi}\right)\, .
\end{align}

More generally, we can establish a matching framework between the 4D
and 5D cosmological quantities that remains valid outside the
inflationary regime. Identifying the line elements of the 5D and 4D
metrics yields the differential relation between the 5D cosmic time
$\hat{t}$ and 4D cosmic time $t$:
\begin{align}
\frac{\partial}{\partial \hat{t}} &= \sqrt{\frac{R}{R_\perp}}\frac{\partial}{\partial t} \, .
\end{align}
This mapping implies a corresponding relation between the 4D and 5D scale factors:
\begin{align} \label{eq:atohata}
a &= \sqrt{\frac{R}{R_\perp}} \hat{a} = \sqrt{\frac{R_0}{R_\perp}} \hat{a}^{3/2}
\end{align}
where we have chosen the normalization $\hat{a}_0 = 1$. The 4D Hubble
parameter $h$ is defined as:
\begin{align}
  h&\equiv \frac{\dot{a}}{a}=\frac{3}{2}\frac{\dot{\hat{a}}}{\hat{a}}=\frac{3}{2}\frac{\dot{R}}{R}\, .
\end{align}
Using the time coordinate transformation, $h$ can be connected to the
5D Hubble parameter $H$ via:
\begin{align} \label{eq:Htoh}
H &\equiv \frac{1}{\hat{a}} \frac{\partial \hat{a}}{\partial \hat{t}}
    = \sqrt{\frac{R}{R_\perp}} \left( h - \frac{1}{2}
    \frac{\dot{R}}{R} \right) = \sqrt{\frac{R}{R_\perp}} \frac{2h}{3}
    \, .
\end{align}

Finally, assuming the energy content of the Universe behaves as a perfect fluid, the 5D and 4D equations of state can be parameterized by $w_{5}$ and $w_{4}$, respectively. Their evolutions are governed by:\begin{align} \label{eq:Hwithw5}
H &= H_i \left( \frac{\hat{a}}{\hat{a}_i} \right)^{-2(1+w_5)}, \qquad h
    = h_i \left( \frac{a}{a_i} \right)^{-\frac{3}{2}(1+w_4)}                            \end{align}
where the subscript $i$ marks the onset of the phase where the equation of state parameter becomes constant. By substituting the scaling relations $R \propto \hat{a} R_0$, \eqref{eq:atohata}, and \eqref{eq:Htoh} into \eqref{eq:Hwithw5}, the 4D Hubble parameter can be rewritten as:
\begin{align}
h &= h_i \left( \frac{a}{a_i} \right)^{-\frac{4}{3}(1+w_5)-\frac{1}{3}}
\end{align}
Equating the exponents yields the final dictionary between the 4D and 5D equation of state parameters:
\begin{align}
w_4 &= \frac{8}{9} w_5 + \frac{1}{9}
\end{align}
This result indicates that the effective 4D equation of state $w_{4}$ is increased compared to its 5D counterpart $w_{5}$ when $w_5\leq1$. This shift is physically expected, as the 4D description explicitly incorporates the kinetic energy density of the evolving radion field.

A summary of the above correspondences is provided in Table~\ref{tab:dict}.

\begin{table}[!h]
    \centering
    \begin{tabular}{c|c|c|c|}
    \cline{2-4}
         & 5D & 4D Jordan & 4D Einstein \\
         \hline
         \multicolumn{1}{|c|}{cosmic time}   & $\hat{t}$ & $\hat{t}$ &$t$\\
        \multicolumn{1}{|c|}{scale factor}   & $\hat{a}$ & $\hat{a}=\frac{R}{R_0}$ &$a=\sqrt{\frac{R}{R_\perp}}\hat{a}=\sqrt{\frac{R_0}{R_\perp}}\hat{a}^{3/2}$\\
        \multicolumn{1}{|c|}{Hubble parameter}   & $H$ & $H$ &$h=\sqrt{\frac{R_\perp}{R}}\frac{3H}{2}$\\
        \multicolumn{1}{|c|}{number of e-folds}   & $\hat{N}$ & $\hat{N}=\ln\left(\frac{R}{R_0}\right)$ &$N=\frac{3\hat{N}}{2}$\\
        \multicolumn{1}{|c|}{equation of state parameter}   & $w_5$ & $w_5$ &$w_4=\frac{8(1+w_5)-7}{9}$\\
        \hline
    \end{tabular}
    \caption{Dictionary between the 5D and 4D frames.}
    \label{tab:dict}
\end{table}

\subsection{Frames, Planck masses, and species scales}\label{sec:C}

In the 4D Einstein frame, we fix the (reduced) Planck mass to the usual $M_p$ by construction:
\begin{align}
    \pi R_\perp M_*^3=M_p^2 \, .
\end{align}
In the Jordan frame, the Planck mass is given by $M_J=\sqrt{\pi R M_*^3}$ and therefore evolves as $R$ grows. In particular, once the extra dimension is stabilized at $R_\perp$, it becomes $M_J(R_\perp)=M_p$ and the two frames are equivalent up to the radion's fluctuations. At the beginning of inflation, however, the Planck mass in the Jordan frame $M_J(R_0)\equiv M_0$ is found to be
\begin{align}
    \pi R_0 M_*^3=M_0^2.
\end{align}

In the presence of a large number of particle species, quantum gravitational effects are expected to become significant at energies well below the Planck scale. This observation motivates the introduction of the species scale \cite{Dvali:2007hz,Dvali:2007wp,vandeHeisteeg:2022btw,Cribiori:2022nke}, which represents the effective ultraviolet cutoff of the theory and is given by\footnote{We introduce a $\pi$ in the definition as we are interested in dimensional reduction. This allows to interpret the species scale as the higher dimensional Planck mass.}
\begin{align}\label{eq:species}
    \Lambda_s=\frac{M}{\sqrt{\pi N_s}}
\end{align}
where  $N_s$ denotes the number of light species below the cutoff and
$M$ denotes the Planck mass (which as previously noted is frame
dependent). Through the explicit dependency in the Planck mass, we
therefore understand that the species scale is also a frame dependent
concept. Let's start by the Jordan frame where the KK mass spectrum is
found to be
\begin{align}
    m^{(J)}_n=\frac{n}{R} \,;
\label{fmJ}
\end{align}
namely, one has $N_{s,J}=R \Lambda_{s,J}$ so that
\begin{align}
    \Lambda_{s,J}=\frac{M_J}{\sqrt{\pi N_{s,J}}}=\sqrt{\frac{\pi R M_*^3}{\pi R \Lambda_{s,J}}}\qquad\qquad\text{giving}\qquad\qquad \Lambda_{s,J}=M_*.
\end{align}
Before proceeding, we pause to note that (\ref{fmJ}) is actually valid for Neumann-Neumann or Dirichlet-Dirichlet boundary conditions, as well as for a compactification on the circle. For Neumann-Dirichlet boundary conditions, one would find a shifted spectrum:
\begin{align}
    m^{(J)}_n=\frac{n+\frac{1}{2}}{R}.
\end{align} 

In the Einstein frame however, the KK spectrum is found to be
\begin{align}\label{eq:KKspectrum}
    m^{(E)}_n=\frac{n}{R}\left(\frac{R_\perp}{R}\right)^{1/2}.
\end{align}
As discussed above, for Neumann-Dirichlet boundary conditions,
the frequencies are half-integers. One can understand (\ref{eq:KKspectrum}) as the masses get rescaled when performing the Weyl rescaling to go into the Einstein frame. This leads to
\begin{align}\label{eq:NE}
    N_{s,E}=R \left(\frac{R}{R_\perp}\right)^{1/2}\Lambda_{s,E}
\end{align}
so that
\begin{align}
    \Lambda_{s,E}=\frac{M_p}{\sqrt{\pi N_{s,E}}}=\frac{M_p}{\sqrt{\pi R \Lambda_{s,E}}}\left(\frac{R_\perp}{R}\right)^{1/4}
\end{align}
giving
\begin{align}
    \Lambda_{s,E}=\left(\frac{M_p^2}{\pi R}\left(\frac{R_\perp}{R}\right)^{1/2}\right)^{1/3}.
\end{align}
Once the radion is stabilized at $R_\perp$, we therefore obtain $\Lambda_{s,E}(R_\perp)=M_*$ and, again, the two frames are equivalent up to the radion's fluctuations. At the beginning of inflation however, one has
\begin{align}\label{eq:Lambda_0}
    \Lambda_0\equiv\Lambda_{s,E}(R_0)=\left(\frac{M_p^2}{\pi R_0}\left(\frac{R_\perp}{R_0}\right)^{1/2}\right)^{1/3}=\left(\frac{R_\perp}{R_0}\right)^{1/2}M_*=\sqrt{\frac{\epsilon}{\pi}}M_p.
\end{align}

At first sight, the Einstein-frame species scale appears to exceed the fundamental five-dimensional scale $M_*$. This is, however, simply a consequence of the Weyl rescaling. Since the Weyl transformation is merely a field redefinition, all quantities with mass dimension one, including the UV cutoff, are rescaled by the same factor. The Einstein-frame species scale should therefore be interpreted as the fundamental five-dimensional cutoff expressed in Einstein-frame units,
\begin{align}
    \Lambda_{s,E}=\left(\frac{R_\perp}{R}\right)^{1/2}M_*.
\end{align}
The apparent enhancement therefore reflects a change of units rather than the emergence of a new physical scale. Consistently, the number of KK states below the cutoff remains unchanged, since it is dimensionless and therefore invariant under the Weyl rescaling. One can indeed check that $N_{s,E}$ given in \eqref{eq:NE} is nothing but $RM_*$, which is also $N_{s,J}$.

In a nutshell, the Jordan frame is the frame where the species scale is fixed while the Planck mass evolves, while the Einstein frame is the frame where we fix the Planck mass and where the effective species scale evolves. These scales are linked through
\begin{align}\label{eq:scalesandframes}
\underset{\text{(5D)}}{\frac{R_\perp}{R_0}}
&= \underset{\text{(J)}}{\left(\frac{M_p}{M_0}\right)^2}
 = \underset{\text{(E)}}{\left(\frac{\Lambda_0}{M_*}\right)^2}
\end{align}
where we have on the left the evolution of the system from the 5D point of view, in the middle from the Jordan frame point of view and on the right from the Einstein frame point of view.

\begin{table}[htbp]
\centering
\begin{tabular}{c | c c|}
\cline{2-3}
 & Planck & Species \\
\hline
\multicolumn{1}{|c|}{Jordan}   & $M_0\rightarrow M_p$ & $M_*$ \\
\multicolumn{1}{|c|}{Einstein} & $M_p$ & $\Lambda_0\rightarrow M_*$ \\
\hline
\end{tabular}
\caption{The Planck mass and the species scale in the Jordan and Einstein frames}
\label{tab:scalesandframes}
\end{table}

A summary of the Planck mass and the species scale in the Jordan and
Einstein frames is given in Table~\ref{tab:scalesandframes}.
\section{Constraining 5D inflation through the TCC}\label{sec:5}

In this section, we turn our attention to the TCC constraints on 5D inflation.

\subsection{The TCC in 5D inflation}
\label{sec:TCC5Dframes}

The TCC for higher dimensional models is addressed in a separate
work~\cite{inpreparation}. In particular, for the 5D case, the TCC is written as\begin{align}\label{eq:5DTCC}
    \frac{\hat{a}_e}{\hat{a}_0}<\frac{M_*}{H_I}.
\end{align}
In the Jordan frame, the species scale $M_*$ and scale factor
$\hat{a}$ yield a constant Hubble parameter $H_{I}$, meaning the
Jordan frame TCC matches the 5D TCC \eqref{eq:5DTCC}.

Conversely, the TCC in the 4D Einstein frame is:
\begin{align}\label{eq:TCCEinstein}
  \frac{a_e}{a_0}<\frac{\Lambda_0}{h_e}.
\end{align}
Using $h_e = 3H_I/2$ and \eqref{eq:Lambda_0}, this becomes:
\begin{align}\label{eq:TCCfactor}
  \frac{a_e}{a_0}\left(\frac{R_0}{R_\perp}\right)^{1/2}<\frac{2}{3}\frac{M_*}{H_I}. \end{align}
Finally, given the relations: \begin{align}
                                \frac{a_e}{a_0}=\left(\frac{R_\perp}{R_0}\right)^{3/2}
                                \quad \text{and} \quad
                                \frac{\hat{a}_e}{\hat{a}_0}=\frac{R_\perp}{R_0}, \end{align}
we see that the Einstein frame TCC \eqref{eq:TCCEinstein} is
equivalent to the Jordan and 5D TCC formulations, differing only by a
factor of $2/3$.

\subsection{Constraint on the inflaton mass}\label{sec:infmass}

In the companion paper~\cite{Anchordoqui:2026los}, we proposed a cosmological history describing the evolution of the Universe from the end of inflation to the reheating epoch, thereby connecting the higher dimensional inflationary phase to the onset of the standard cosmological evolution. As discussed in Sec.~\ref{sec:4Dinf}, the presence of a large extra dimension constrains the reheating temperature to lie below the so-called normalcy temperature, $T_r \leq T_N \sim 1\,\text{GeV}$, in order to avoid the overproduction of KK gravitons, which would otherwise overclose the Universe~\cite{Arkani-Hamed:1998sfv}.

The post-inflationary evolution prior to reheating is described by two distinct phases. The first is an intermediate epoch characterized by a model dependent effective equation-of-state parameter $w$. This is followed by a phase in which the inflaton oscillates around the minimum of its potential once the Hubble parameter drops below the inflaton mass $m$. Assuming that the potential is approximately quadratic near its minimum, these coherent oscillations behave as pressureless matter, corresponding to an effective equation-of-state parameter $w_{ os}=0$. The successive cosmological phases are summarized in Table~\ref{tab:phases1}.

\begin{table}[h]
\centering
\renewcommand{\arraystretch}{1.5} 

\begin{tabular}{r|c|c|c|l}
\cline{2-4}
\smash{\raisebox{-12pt}{beginning of inflation $\rightarrow$}}& $w_5$ & $w_4$ & $N$ & \smash{\raisebox{-12pt}{$\leftarrow$ $R_0$}}\\ 
\hhline{~===~}

\smash{\raisebox{-10pt}{end of inflation $\rightarrow$}}& $-1$ & $-7/9$ & $N$& \smash{\raisebox{-10pt}{$\leftarrow$ $R_\perp$}} \\ 
\cline{2-4}
\smash{\raisebox{-10pt}{beginning of oscillations $\rightarrow$}}&  & $w$ & $\Delta N$ & \\
\cline{2-4}
\smash{\raisebox{-10pt}{reheating $\rightarrow$}}&  & $0$ & $N_{os}$ &  \\
\cline{2-4}
\smash{\raisebox{-10pt}{today $\rightarrow$}}&  & $\sim1/3$ & $N_r$ &  \\
\cline{2-4}

\end{tabular}
\caption{The different phases of the cosmological evolution in Sec.~\ref{sec:infmass}}
\label{tab:phases1}
\end{table}

The parameter space of the model is primarily constrained by two requirements: the generation of a nearly scale invariant primordial power spectrum, consistent with CMB observations, and the resolution of the horizon problem, thereby accounting for the observed homogeneity of the Universe on large scales.

As mentioned in the Introduction, higher dimensional inflation predicts a primordial power spectrum that departs from scale invariance on the largest scales, corresponding to fluctuations whose initial wavelengths exceed the compactification scale~\cite{Anchordoqui:2023etp}. On smaller scales, the summation over the KK modes restores the standard nearly scale invariant spectrum. More precisely, the angular power spectrum of scalar density perturbations can be expressed in terms of the slow-roll parameters and is given by\footnote{The overall normalization differs from that adopted in Refs.~\cite{Antoniadis:2023sya,Antoniadis:2025pet} due to the choice of mass units, with $M_{0}^2 \equiv 2\pi R_0 M_*^3$, where $M_{0}$ denotes the reduced Planck mass at the onset of inflation. This relation holds for a circular extra dimension. For an interval, the factor of $2$ is absent because the volume is simply $\pi R_0$. However, the KK summation acquires an additional factor of $1/2$, since only even KK modes are present, leaving the final result unchanged.}~\cite{Antoniadis:2023sya,Antoniadis:2025pet,Hirose:2025pzm}
\begin{align}\label{eq:pre_37}
    P_\mathcal{R} &\underset{R_0k \ll 1}{\simeq} \frac{H_I^3}{3\pi^4\varepsilon R_0kM_*^3}\biggl[\biggl(\frac{k}{\hat{a}H_I}\biggr)^{2\delta - 5\varepsilon}+\frac{\varepsilon}{3}\biggl(\frac{k}{\hat{a}H_I}\biggr)^{-3\varepsilon}\biggr],\\
    \label{eq:pre_47}
    P_\mathcal{R} &\underset{R_0k \gg 1}{\simeq} \frac{H_I^3}{6\pi^3\varepsilon M_*^3}\biggl[\biggl(\frac{k}{\hat{a}H_I}\biggr)^{2\delta - 5\varepsilon}+\frac{5\varepsilon}{24}\biggl(\frac{k}{\hat{a}H_I}\biggr)^{- 3\varepsilon}\biggr],
\end{align}
whereas the power spectrum of the tensor modes is found to be
\begin{eqnarray}\label{eq:pre_48}
    P_\mathcal{T} \underset{R_0k \ll 1}{\simeq} \frac{8H_I^3}{\pi^4 R_0kM_*^3}\biggl(\frac{k}{\hat{a}H_I}\biggr)^{- 3\varepsilon}
    \quad\text{and}\quad
    P_\mathcal{T} \underset{R_0k \gg 1}{\simeq}
  \frac{4H_I^3}{\pi^3M_*^3}\biggl(\frac{k}{\hat{a}H_I}\biggr)^{-
  3\varepsilon} \,.
\end{eqnarray}
The slow-roll parameters are defined through the Hubble flow hierarchy as $\varepsilon = - \hat{\dot H}_I/H_I^2\equiv\varepsilon_1$ and
$\varepsilon_{n+1}={\hat{\dot\varepsilon}_n/(H_I\varepsilon_n)}$, together with $\delta
= \varepsilon-\varepsilon_2/2$, where a hatted dot denotes differentiation with respect to $\hat{t}$. The tensor-to-scalar ratio is then given by $r=24\varepsilon$. In the range of scales probed by the {\it Planck} satellite, corresponding to $R_0k\gg1$, the KK summation restores the standard nearly scale invariant spectrum, and the tensor-to-scalar ratio can be expressed, at leading order in the slow-roll expansion, as
\begin{align}
    r=\frac{4H_I^3}{\pi^3A_sM_*^3}=\frac{4H_I^3R_\perp}{\pi^2A_sM_p^2}=1.6\times10^{-19} \left(\frac{H_I}{\text{GeV}}\right)^3 \left(\frac{R_\perp}{\mu\text{m}}\right) \, ,
\label{eq:r}
\end{align}
where $A_s$ is the amplitude of the adiabatic scalar perturbations measured by {\it Planck}, $A_s\sim2.1\times10^{-9}$~\cite{Planck:2018vyg}. In order to be compatible with {\it Planck} data, the transition must occur at multipoles below $\ell\simeq3.23$, so that the violation of scale invariance is essentially absent from the observable CMB spectrum~\cite{Anchordoqui:2026los,Petretti:2024mjy}. This, in turn, constrains the equation-of-state parameter to the range
$-1/3 \leq w \lesssim 0$. In particular, the allowed region of parameter space is maximized for $w=-1/3$. 
This equation of state is obtained effectively in the absence of matter by the linear dilaton background which is an exact string solution and is discussed later, in Sec.~\ref{subsec:leU}, thus providing an independent theoretical motivation of such a phase. Moreover, within M-theory the radion of the 11th dimension is identified with the dilaton of the heterotic string.

In the following, we therefore adopt the value of the linear expansion ($w=-1/3$ in 4D) in order to determine the maximal allowed parameter space of the model. 
The requirement of scale invariance finally yields~\cite{Anchordoqui:2026los}
\begin{align}\label{eq:SIw=-1/3}
    \left(\frac{m}{\text{GeV}}\right)^{1/3}\leq1.5\times10^{-6}g_*^{-1/3}\left(\frac{T_r}{\text{GeV}}\right)^{-1/3}\left(\frac{R_\perp}{\mu\text{m}}\right)\left(\frac{H_I}{\text{GeV}}\right)
\end{align}
while the horizon problem constraint becomes independent of $H_I$:
\begin{align}\label{eq:horw=-1/3}
    \left(\frac{m}{\text{GeV}}\right)^{1/3}\leq70\,\epsilon\, g_*^{-1/3}\left(\frac{T_r}{\text{GeV}}\right)^{-1/3}\left(\frac{R_\perp}{\mu\text{m}}\right)^{2/3}.
\end{align}
Combining these expressions, one finds
\begin{align}\label{eq:m}
    \left(\frac{m}{\text{GeV}}\right)^{2/3}\leq1.0\times10^{-4}\,\epsilon\, g_*^{-2/3}\left(\frac{T_r}{\text{GeV}}\right)^{-2/3}\left(\frac{R_\perp}{\mu\text{m}}\right)^{5/3}\left(\frac{H_I}{\text{GeV}}\right).
\end{align}

Finally, we turn to the TCC. In the Einstein frame, the TCC constraint \eqref{eq:TCCEinstein} can be rewritten as
\begin{align}
    e^N<\frac{2M_p}{3H_I}\sqrt{\frac{\epsilon}{\pi}}\qquad\qquad\text{giving}\qquad\qquad H_I<\frac{2}{3\epsilon R_\perp},
\label{TCCcon}
\end{align}
leading to
\begin{align}\label{eq:TCCEunits}
    \frac{H_I}{\text{GeV}}<1.3\times10^{-10}\,\epsilon^{-1}\left(\frac{R_\perp}{\mu\text{m}}\right)^{-1}.
\end{align}
Together with \eqref{eq:m}, it implies
\begin{align}
   \frac{m}{\text{GeV}}\leq1.6\times10^{-21}\, g_*^{-1}\left(\frac{T_r}{\text{GeV}}\right)^{-1}\left(\frac{R_\perp}{\mu\text{m}}\right).
\end{align}

The size of the extra dimension is constrained by short-range tests of Newton's inverse-square law, which require $R_\perp\lesssim40\,\mu\text{m}$ in the case of an interval (or $R_\perp\lesssim30\,\mu\text{m}$ for a circle)~\cite{Lee:2020zjt,Tan:2020vpf}.\footnote{Short-range gravity tests were first proposed to distinguish between these two geometries in~\cite{Schwarz:2024tet}, with current bounds derived in~\cite{Anchordoqui:2026hys}.} We therefore consider the most favorable allowed value, $R_\perp=40\,\mu\text{m}$. Together with the lowest reheating temperature compatible with BBN, $T_r=5\,\text{MeV}\gtrsim T_{\rm BBN}$, corresponding to $g_*=10.74$~\cite{Husdal:2016haj}, this yields the upper bound $m\lesssim10^{-9}\,\text{eV}$. Such an extremely small inflaton mass is difficult to reconcile with the post-inflationary dynamics assumed throughout this work, namely coherent oscillations in an approximately quadratic potential. We therefore conclude that, within the framework considered here, the simultaneous implementation of the TCC and the cosmological constraints leads to an implausible parameter regime, suggesting that this simple post-inflationary scenario is strongly disfavored by TCC.

\subsection{Extending the 5D cosmological evolution}\label{sec:5Dalpha}
Since the TCC is compatible only with phenomenologically unacceptable values of $m$, it effectively requires the intermediate phase with $w=-1/3$ to persist until reheating, leaving no room for a subsequent phase of inflaton oscillations. We therefore discard the oscillatory phase and instead consider a cosmological history in which reheating occurs directly from the $w=-1/3$ phase. Possible reheating mechanisms for a runaway inflaton have been discussed in \cite{Felder:1999pv}.

We further generalize our analysis by allowing the extra dimension to continue expanding after the end of inflation before stabilizing it at a later time. As before, we consider the most favorable scenarios, corresponding to phases of linear expansion, first in the five-dimensional regime and subsequently after the transition to the four-dimensional regime. The resulting cosmological histories, consisting of four successive phases, are summarized in Table~\ref{tab:phases2}.

\begin{table}[h]
\centering
\renewcommand{\arraystretch}{1.5} 

\begin{tabular}{r|c|c|c|l}
\cline{2-4}
\smash{\raisebox{-12pt}{beginning of inflation $\rightarrow$}}& $w_5$ & $w_4$ & $N$ & \smash{\raisebox{-12pt}{$\leftarrow$ $R_0$}}\\ 
\hhline{~===~}

\smash{\raisebox{-10pt}{end of inflation $\rightarrow$}}& $-1$ & $-7/9$ & $N$& \smash{\raisebox{-10pt}{$\leftarrow$ $R_e$}} \\ 
\cline{2-4}
\smash{\raisebox{-10pt}{stabilization of $R$ $\rightarrow$}}& $-1/2$ & $-1/3$ & $N_5$ & \smash{\raisebox{-10pt}{$\leftarrow$ $R_\perp$}} \\
\cline{2-4}
\smash{\raisebox{-10pt}{reheating $\rightarrow$}}&  & $-1/3$ & $N_4$ &  \\
\cline{2-4}
\smash{\raisebox{-10pt}{today $\rightarrow$}}&  & $\sim1/3$ & $N_r$ &  \\
\cline{2-4}

\end{tabular}
\caption{The different phases of the cosmological evolution in Sec.~\ref{sec:5Dalpha}}
\label{tab:phases2}
\end{table}

The number of e-folds of each phase can be found to be
\begin{align}
    e^{N}&=\frac{a_e}{a_0}=\left(\frac{h_e}{h_0}\right)^{-3}=\left(\frac{R_e}{R_0}\right)^{3/2}=R_\perp M_p\frac{(\alpha\epsilon)^{3/2}}{\sqrt{\pi}},\label{eq:eN}\\
    e^{N_5}&=\frac{a_\perp}{a_e}=\left(\frac{h_\perp}{h_e}\right)^{-1}=\left(\frac{R_\perp}{R_e}\right)^{3/2}=\alpha^{-3/2},\\
    e^{N_4}&=\frac{a_r}{a_\perp}=\left(\frac{h_r}{h_\perp}\right)^{-1}=\left(\sqrt{\frac{\pi^2g_*}{90}}\frac{T_r^2}{M_p}\frac{2}{3H_I}\alpha^{-1}\right)^{-1},\label{eq:N4}
\end{align}
altogether with \eqref{eq:N_r} and where we introduced the parameter $\alpha=R_e/R_\perp\leq 1$. The fact that the number of e-folds during the 5D phases is simply given by the ratios of the final radius over the initial radius to the power $3/2$, independently of the value of $w_5$, is simply the consequence of the fact that $a\propto\hat{a}^{3/2}\propto R^{3/2}$ while the Hubble parameter at the time of the stabilization of the extra dimension $h_\perp$ can be found as
\begin{align}
    h_\perp=\frac{3H_I}{2}\frac{R_e}{R_\perp}
\end{align}
using \eqref{eq:Htoh} and \eqref{eq:Hwithw5}. Requiring $N$, $N_5$, and $N_4$ to be positive further implies the constraints
\begin{align}
    R_0\leq R_e\leq R_\perp\qquad\qquad\text{and}\qquad\qquad h_\perp\geq h_r.
  \end{align}

We now reconsider the scale invariance condition mentioned in the previous subsection. In the present cosmological scenario, the scale invariance requirement $\pi kR_0\geq1$ can be rewritten as
\begin{align}
    \frac{2\pi^2}{\lambda_t}\frac{a_t}{a_r}\frac{a_r}{a_\perp}a_\perp R_0\geq1\qquad\text{giving}\qquad
2\pi^2 R_\perp e^{N_4+N_r}\geq\lambda_t
\end{align}
where $\lambda_t\geq27.3\,\text{Gpc}$ is the transition wavelength today~\cite{Anchordoqui:2026los,Petretti:2024mjy}. It can further be expressed as
\begin{align}\label{eq:SIalpha}
    1\geq\alpha\geq 3.5\times10^{-1}\,g_*^{1/2}\left(\frac{T_r}{\text{GeV}}\right)\left(\frac{H_I}{\text{GeV}}\right)^{-1}\left(\frac{R_\perp}{\mu\text{m}}\right)^{-1}.
\end{align}

On the other hand, the condition for solving the horizon problem reads
\begin{align}\label{eq:hor522}
    D_{\text{LSS}}\leq\frac{a_t}{a_0h_0}= e^{N+N_5+N_4+N_r}\frac{1}{h_0}=e^{\frac{2}{3}N+\frac{2}{3}N_5+N_4+N_r}\frac{2}{3H_I}
\end{align}
where we used $h_0=(3H_I/2)(R_\perp/R_0)^{1/2}$, leading to
\begin{align}\label{eq:horalpha}
    1\geq\alpha\geq7.4\times10^{-9}\,g_*^{1/2}\,\epsilon^{-1}\left(\frac{T_r}{\text{GeV}}\right)\left(\frac{R_\perp}{\mu\text{m}}\right)^{-2/3}.
\end{align}

The final constraint is provided by the TCC. In the 4D Einstein frame, the TCC condition~\eqref{eq:TCCEinstein} can be expressed as
\begin{align}\label{eq:TCCalpha}
    \alpha\leq1.3\times10^{-10}\,\epsilon^{-1}\left(\frac{H_I}{\text{GeV}}\right)^{-1}\left(\frac{R_\perp}{\mu\text{m}}\right)^{-1}
\end{align}
where we used $h_e=3\alpha^{-1/2}H_I/2$.

\begin{figure}[h]
     \begin{subfigure}[t]{0.5\textwidth}
    \centering\includegraphics[width=0.98\linewidth]{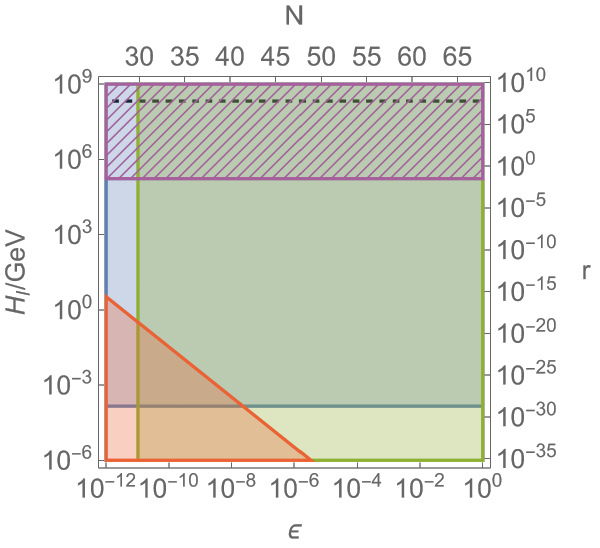}
    \subcaption{$R_\perp=40\,\mu\text{m},T_r=5\,\text{MeV},\,\alpha=1$}
    \label{fig:alpha1}
    \end{subfigure}
    \begin{subfigure}[t]{0.5\textwidth}
    \centering\includegraphics[width=0.98\linewidth]{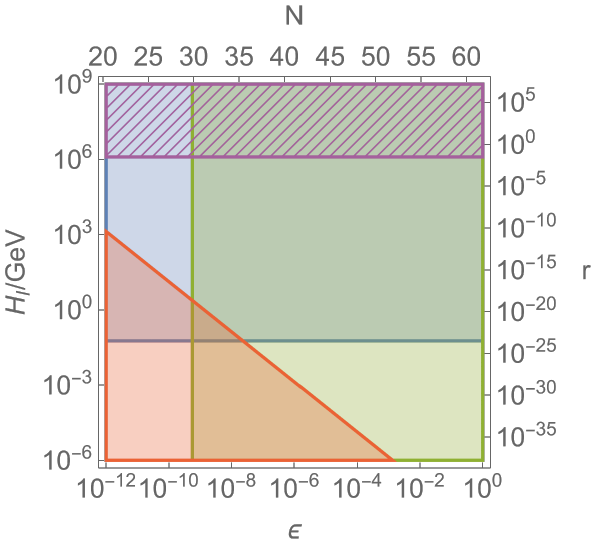}
    \subcaption{$R_\perp=0.1\,\mu\text{m},T_r=5\,\text{MeV},\,\alpha=1$}
    \label{fig:alpha2}
    \end{subfigure}

\vspace{0.3cm}
     \begin{subfigure}[t]{0.5\textwidth}
    \centering\includegraphics[width=0.98\linewidth]{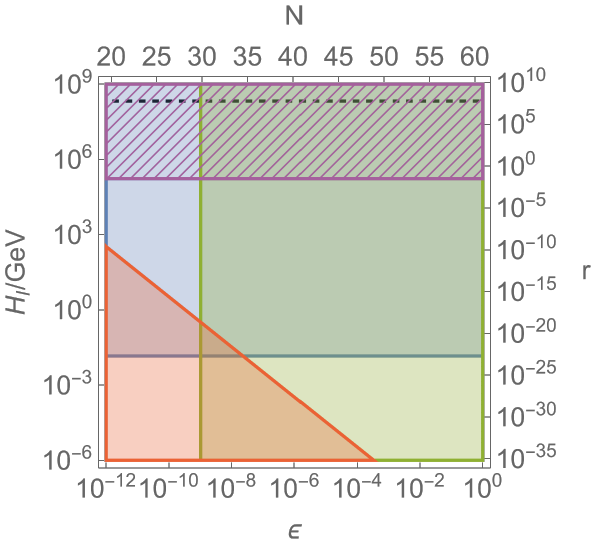}
    \subcaption{$R_\perp=40\,\mu\text{m},T_r=5\,\text{MeV},\,\alpha=10^{-2}$}
    \label{fig:alpha3}
    \end{subfigure}
    \begin{subfigure}[t]{0.5\textwidth}
    \centering\includegraphics[width=0.98\linewidth]{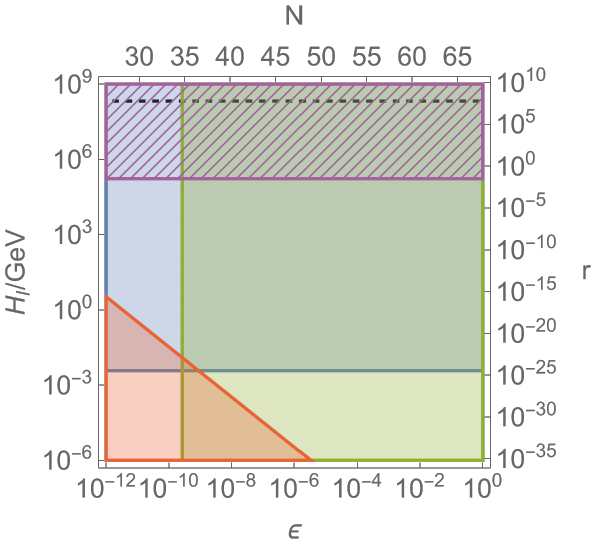}
    \subcaption{$R_\perp=40\,\mu\text{m},T_r=100\,\text{MeV},\,\alpha=1$}
    \label{fig:alpha4}
    \end{subfigure}

  \caption{Constraints on 5D inflation. The blue, green and red regions correspond to the regions allowed by the scale invariant condition \eqref{eq:SIalpha}, by the horizon problem constraint \eqref{eq:horalpha} and by the TCC  \eqref{eq:TCCalpha}, respectively. Thus, the allowed region of the parameter space corresponds to the intersection of the three. The hatched region corresponds to $r > 0.032$ excluded by experiments (derived using a combination of BICEP/Keck 2018 and Planck data~\cite{BICEP:2021xfz,Tristram:2021tvh}) and the black dashed line corresponds to the value of $M_*$. We use $g_*=10.74$ for $T_r=5\,\text{MeV}$ and $g_*\sim18.00$ for $T_r=100\,\text{MeV}$ \cite{Husdal:2016haj}. The corresponding values of  the tensor-to-scalar ratio $r$ and of the number of e-folds $N$ are indicated on the right side and on the top, respectively.}
  \label{fig:alpha}
\end{figure}

We now turn to determining the allowed region of parameter space. The parameter $\epsilon$ is constrained by the scale invariance condition~\eqref{eq:SIalpha} together with the TCC bound~\eqref{eq:TCCalpha}, from which we obtain
\begin{align}
    \epsilon\leq3.8\times10^{-10}\,g_*^{-1/2}\left(\frac{T_r}{\text{GeV}}\right)^{-1}\lesssim 2.3\times10^{-8}
\end{align}
where the upper bound is obtained by taking $T_r=5\,\mathrm{MeV}\gtrsim T_{\rm BBN}$, corresponding to $g_*=10.74$~\cite{Husdal:2016haj}. Similarly, combining the horizon problem condition~\eqref{eq:horalpha} with the TCC bound~\eqref{eq:TCCalpha} yields the following constraint on $H_I$:
\begin{align}\label{eq:maxHI}
    \left(\frac{H_I}{\text{GeV}}\right)\leq1.8\times10^{-2}\,g_*^{-1/2}\left(\frac{T_r}{\text{GeV}}\right)^{-1}\left(\frac{R_\perp}{\mu\text{m}}\right)^{-1/3}\lesssim1.1\left(\frac{R_\perp}{\mu\text{m}}\right)^{-1/3}
\end{align}
where the upper bound is derived, once again, by taking
$T_r=5\,\mathrm{MeV}\gtrsim T_{\rm BBN}$. Substituting this bound into \eqref{eq:r} gives
\begin{align}
    r\lesssim2.2\times10^{-19}.
\end{align}
The TCC therefore imposes stringent constraints on the model parameters that are independent of $\alpha$, as illustrated in Fig.~\ref{fig:alpha}. As a consequence, the amplitude of primordial gravitational waves is strongly suppressed. Nevertheless, the resulting upper bound on the tensor-to-scalar ratio remains significantly weaker than the one obtained in the case of 4D inflation with a stabilized large extra dimension, \eqref{eq:r4D}.

We conclude this analysis by noting that, instead of a subsequent phase of 5D evolution, one could envisage a phase of 4D inflation immediately following the end of 5D inflation. Such a modification, however, does not relax the constraints. Indeed, at the end of the five-dimensional inflationary phase one has $h_e=3H_I/2$, so that the Higuchi bound requires $2h_e^2=9H_I^2/2\leq1/R_\perp^2$. This again leads, up to a factor of order unity, to the same constraints as those derived in Sec.~\ref{sec:4Dinf}.

\section{A pre-inflationary history of the Universe}\label{sec:6}

The constraints imposed by the TCC can be summarized as follows: \emph{(i)} the extra dimension must already be hierarchically larger than the fundamental scale at the onset of inflation, $\epsilon\lesssim10^{-8}$, and \emph{(ii)} the Hubble radius must likewise be hierarchically larger than the fundamental scale since \eqref{eq:maxHI} can be rewritten as $H_I/M_*\lesssim10^{-9}$. While the first condition provides \emph{a posteriori} justification for describing the inflationary dynamics within a five-dimensional effective field theory, the two conditions taken together may instead indicate that the Universe underwent a non-negligible pre-inflationary expansion, during which both $R^{-1}$ and $H$ decreased from values of order $M_*$ to the initial values assumed at the onset of inflation.

\begin{figure}[h]
     \begin{subfigure}[t]{0.5\textwidth}
    \centering\includegraphics[width=0.98\linewidth]{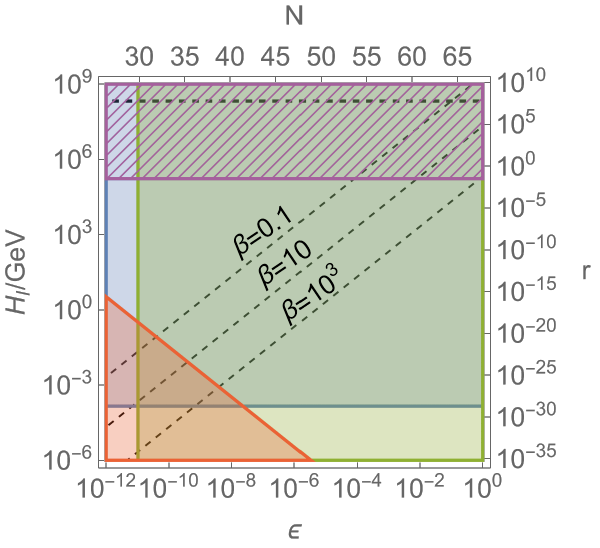}
    \subcaption{$R_\perp=40\,\mu\text{m},T_r=5\,\text{MeV},\,\alpha=1$}
    \label{fig:beta1}
    \end{subfigure}
    \begin{subfigure}[t]{0.5\textwidth}
    \centering\includegraphics[width=0.98\linewidth]{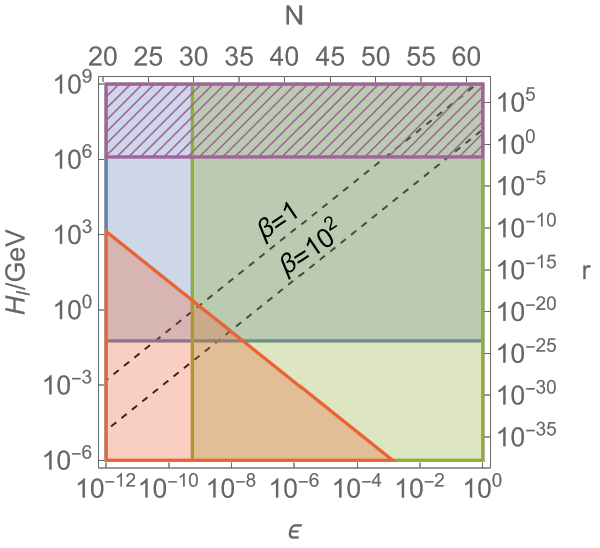}
    \subcaption{$R_\perp=0.1\,\mu\text{m},T_r=5\,\text{MeV},\,\alpha=1$}
    \label{fig:beta2}
    \end{subfigure}

  \caption{The figures Fig.~\ref{fig:alpha1} and Fig.~\ref{fig:alpha2} where the solutions of \eqref{eq:beta} are shown for different values of $\beta$.}
  \label{fig:beta}
\end{figure}

Let us therefore consider such a pre-inflationary phase, during which the initial values $R_i$ and $H_i$ evolve to $R_0$ and $H_I$ at the onset of inflation. Assuming that this phase is characterized by an equation-of-state parameter $w_5$, we have
\begin{align}\label{eq:HItoHi}
    H_I=H_i\left(\frac{R_0}{R_i}\right)^{-2(1+w_5)}.
\end{align}
In particular, for a phase of linear expansion, corresponding to $w_5=-1/2$, and assuming the `natural' initial condition $R_i^{-1}=\beta H_i$, where $H_i$ could be as large as $M_*$, \eqref{eq:HItoHi} reduces to $H_I=(\epsilon/\beta)M_*$, or equivalently,
\begin{align}\label{eq:beta}
    \left(\frac{H_I}{\text{GeV}}\right)=7.2\times10^{8}\,\frac{\epsilon}{\beta}\left(\frac{R_\perp}{\mu\text{m}}\right)^{-1/3}.
\end{align}
Moreover, the simultaneous saturation of the scale invariance condition~\eqref{eq:SIalpha} and the horizon problem constraint~\eqref{eq:horalpha}, corresponding to the intersection of the solid green and blue lines in Fig.~\ref{fig:alpha}, yields
\begin{align}
    \left(\frac{H_I}{\text{GeV}}\right)=4.8\times10^{7}\,\epsilon\left(\frac{R_\perp}{\mu\text{m}}\right)^{-1/3}.
\end{align}
Thus, $\beta\sim10$ is sufficient to ensure natural initial conditions whenever the TCC constraint~\eqref{eq:TCCalpha} intersects the scale invariance constraint~\eqref{eq:SIalpha} and the horizon problem constraint~\eqref{eq:horalpha}. Other values of $\beta$ are also compatible with natural initial conditions, provided that the allowed region of parameter space does not collapse to a single point; see, for instance, Fig.~\ref{fig:beta}. Moreover, the number of e-folds in the 4D Einstein frame $N_i$ during this pre-inflationary phase is given by\footnote{This assumes that the dynamics is purely 5D, corresponding to a 5D number of e-folds $\hat{N}_i=2N_i/3\sim18$. This phase could also encompass an earlier higher-dimensional evolution responsible for the stabilization of the remaining extra dimensions. See for instance Sec.~\ref{sec:Mtheory}.
}
\begin{align}
    e^{N_i}=\left(\frac{R_0}{R_i}\right)^{3/2}=(\epsilon R_i M_*)^{-3/2}=\left(\frac{\beta}{\epsilon}\frac{H_i}{M_*}\right)^{3/2}.
\end{align}
For natural values $\beta\sim1$ and $H_i/M_*\sim1$, one therefore finds $N_i\sim27$ in order to obtain the required value $\epsilon\sim10^{-8}$.

One may wonder why we have restricted our analysis to the case $w_5=-1/2$. More generally, smaller values of $w_5$ could equally well be considered. Our purpose in this section, however, is simply to illustrate the role of a pre-inflationary phase in a model-independent way, while leaving the discussion of more complete and realistic scenarios to Sec.~\ref{sec:Mtheory}. In particular, the choice $w_5=-1/2$, corresponding to a phase of linear expansion, arises naturally in the so-called linear dilaton solution, providing a well-motivated realization of the cosmological history considered here.

Considering a pre-inflationary phase also modifies the condition required to solve the horizon problem. Indeed, in the absence of such a phase, the observable Universe must originate from a single Hubble patch at the onset of inflation, leading to the usual horizon problem constraint. Once a pre-inflationary era is included, however, causal contact can already be established before inflation. The relevant quantity is therefore the particle horizon rather than the Hubble radius. The horizon problem is then solved provided the physical size of the last scattering surface satisfies
\begin{align}
    \frac{D_{\rm LSS}}{a_t}\leq \chi(t_{\rm LSS}),
\end{align}
where
\begin{align}
    \chi(t_\text{LSS})=\int_{t_i}^{t_\text{LSS}}\frac{dt}{a(t)}=\int_{\hat{t}_i}^{\hat{t}_\text{LSS}}\frac{d\hat{t}}{\hat{a}(\hat{t})}=\hat{\chi}(\hat{t}_{\text{LSS}})
\end{align}
is the comoving particle horizon evaluated at the time of last scattering, with the integral receiving contributions from both the pre-inflationary phase and the subsequent inflationary evolution. In particular, the comoving particle horizon is frame independent and is therefore given by the same expression whether computed in 4D or in 5D. The comoving particle horizon can further be split into the contributions from each cosmological phase by decomposing the integral as
\begin{align}
    \chi(t_{\rm LSS})
    =\int_{\hat t_i}^{\hat t_0}\frac{d\hat t}{\hat a(\hat t)}
    +\int_{\hat t_0}^{\hat t_e}\frac{d\hat t}{\hat a(\hat t)}
    +\cdots
    \equiv
    \chi_i+\chi_I+\cdots,
\end{align}
where the different terms correspond to the successive stages of the cosmological evolution. The contribution from inflation, extending from $\hat t_0=0$ to $\hat t_e$, with $\hat a(\hat t)=e^{H_I\hat t}$,
is given by
\begin{align}
    \chi_I
    =
    \int_{\hat t_0}^{\hat t_e}\frac{d\hat t}{\hat a(\hat t)}
    =
    \frac{1-e^{-\hat N}}{H_I},
\end{align}
which, for $\hat N\gg1$, is dominated by
\begin{align}
    \chi_I\simeq\frac{1}{H_I}
    =\frac{1}{\hat a_0H_I}
    =\frac{3}{2a_0h_0}.
\end{align}
This explains why, in the absence of any earlier cosmological phase, the horizon problem is usually formulated in terms of the initial Hubble radius, requiring\footnote{We note that this derivation yields an additional factor of $3/2$ compared to the naive expression \eqref{eq:hor522}, which is written directly in terms of the 4D Einstein frame Hubble parameter $h_0$.}
\begin{align}
    \frac{D_{\rm LSS}}{a_t}\lesssim\frac{3}{2a_0h_0}.
\end{align}
We note that the phases following inflation are irrelevant, as their contributions to the comoving particle horizon are exponentially suppressed by the inflationary expansion. For instance, a 4D linear phase of $N_4$ e-folds directly following inflation (namely $\alpha=0$) up to reheating, extending from $t_e$ to $t_r$, is described by
\begin{align}
    a(t)=e^{\hat N}\left(1+\frac{3H_I}{2}(t-t_e)\right)
         =e^{\hat N}\left(\frac{3H_I}{2}t-2\right),
\end{align}
where we used $t_e=2/H_I$, so that $a(t_e)=a_e=e^{\hat N}$ and $h(t_e)=h_e=3H_I/2$. Its contribution to the particle horizon is
\begin{align}
    \chi_4=\int_{t_e}^{t_r}\frac{dt}{a(t)}
    =\frac{2N_4}{3H_I}e^{-\hat N},
\end{align}
which is exponentially suppressed for $\hat N\gg1$. Similarly, for $\alpha>1$, the phase immediately following inflation is a 5D linear expansion of $N_5$ e-folds in the 4D Einstein frame, with
\begin{align}
    \hat a(\hat t)
    =\hat a_e\left(1+H_I(\hat t-\hat t_e)\right)
    =e^{\hat N}\left(H_I\hat t+1-\hat N\right),
\end{align}
leading to
\begin{align}
    \chi_5
    =\int_{\hat t_e}^{\hat t_\perp}\frac{d\hat t}{\hat a(\hat t)}
    =\frac{2N_5}{3H_I}e^{-\hat N}.
\end{align}
The contributions from all subsequent cosmological phases are suppressed by the same exponential factor, or an even stronger one, and are therefore negligible.

In the absence of any pre-inflationary evolution, the initial Hubble radius provides therefore an excellent approximation to the particle horizon. As we now show, however, a pre-inflationary expansion can give a sizeable contribution to the latter and consequently modify the horizon problem condition. Indeed, let us consider a 5D linear pre-inflationary expansion extending from $\hat{t}_i$ to the onset of inflation $\hat{t}_0$, described by
\begin{align}
    \hat{a}(\hat{t})=1+H_I\hat{t}.
\end{align}
Its contribution to the comoving particle horizon is then given by
\begin{align}
    \chi_i
    =\int_{\hat t_i}^{\hat t_0}\frac{d\hat t}{\hat a(\hat t)}
    =\frac{\hat N_i}{H_I},
\end{align}
where
\begin{align}
    \hat N_i
    =\ln\left(\frac{H_i}{H_I}\right)
    =\ln\left(\frac{R_0}{R_i}\right)
\end{align}
denotes the number of e-folds of the pre-inflationary expansion. In particular, considering $R_i\sim M_*^{-1}$ gives $\hat N_i\simeq-\ln(\epsilon)$. The particle horizon is therefore enhanced by the pre-inflationary contribution, and solving the horizon problem requires
\begin{align}
    \frac{D_{\rm LSS}}{a_t}
    \lesssim
    \frac{1-\ln(\epsilon)}{H_I}.
\end{align}
Therefore, \eqref{eq:horalpha} is modified into
\begin{align}\label{eq:horalpha2}
    1\geq\alpha\geq4.9\times10^{-9}\,g_*^{1/2}\,\frac{\epsilon^{-1}}{1-\ln(\epsilon)}\left(\frac{T_r}{\text{GeV}}\right)\left(\frac{R_\perp}{\mu\text{m}}\right)^{-2/3}
\end{align}
which slightly extends the green regions towards smaller values of $\epsilon$ in the plots, thereby mildly relaxing the upper bound on the inflationary scale,
$H_I\lesssim10^2-10^3\,\text{GeV}$, and thus the upper bound of the ratio of tensor-to-scalar primordial perturbations to $r\lesssim 10^{-10}$ (see~\eqref{eq:r}). On the other hand, analysing the impact of the pre-inflationary phase on the inflaton oscillation scenario discussed in Sec.~\ref{sec:infmass}, we find that it only increases the allowed inflaton mass by about two orders of magnitude, up to
$m\lesssim10^{-7}\,\text{eV}$. Such a low mass remains difficult to reconcile with the post-inflationary dynamics assumed in~\cite{Anchordoqui:2026los}, which relies on coherent inflaton oscillations in an approximately quadratic potential.

We end up this section by mentioning that such pre-inflationary expansion could also be considered for the case of 4D inflation in the presence of a stabilized extra dimension studied in Sec.~\ref{sec:4Dinf}, in order to address the extremely small Hubble scale required for consistency with the Higuchi bound. In this case we have
\begin{align}
    H_I=\frac{3H_i}{2}\left(\frac{R_\perp}{R_i}\right)^{-2(1+w_5)}.
\end{align}
Considering again a phase of linear expansion, $w_5=-1/2$, together with the initial condition $R_i^{-1}=\beta H_i$, \eqref{eq:HItoHi} reduces to $H_I=3/(2\beta R_\perp)$, which naturally satisfies the Higuchi bound for $\beta\sim1$. In this scenario, the number of e-folds in the 4D Einstein frame is given by
\begin{align}
e^{N_i}=\left(\frac{R_\perp}{R_i}\right)^{3/2}=(\beta H_iR_\perp)^{3/2},
\end{align}
so that starting from $R_i^{-1}\sim M_*$ and ending with $R_\perp$ of order the micron requires $N_i\sim60$, corresponding to $\hat{N}_i=2N_i/3\sim40$ e-folds of 5D expansion. In this picture, the stabilization of the radion at the end of the 5D expansion could naturally provide the trigger for this subsequent 4D inflationary phase.

Interestingly, a related cosmological sequence has recently been proposed in \cite{Yamada:2026ipx}, where curvature-assisted dynamical compactification drives the radion toward a stabilized vacuum before the onset of 4D inflation. In the curvature-dominated regime, the effective four-dimensional equation of state is $w_4=-1/3$, corresponding to a linear expansion phase, thus providing a concrete realization of the type of pre-inflationary evolution considered here.

\section{Constraining 6D inflation through the TCC}\label{sec:7}

As already mentioned in the Introduction, the case of two extra dimensions remains marginally viable. We therefore turn to the study of the TCC in this context.

\subsection{The TCC in arbitrary higher dimensional inflation}

While we showed in Sec.~\ref{sec:TCC5Dframes} that the TCC is frame independent, we start by showing here that it also leads to the same constraint, independently of the number of dimensions, in the case of uniform inflation. Indeed, in $(4+d)$-dimensional uniform inflation, the scale factor is directly related to the size of the extra dimensions, $\hat{a}\sim R$. Therefore, expressing the TCC in terms of the species scale $M_*$, defined by $(\pi R_\perp)^d M_*^{2+d}=M_p^2$, gives
\begin{align}
    \frac{\hat{a}_e}{\hat{a_0}}=\frac{R_\perp}{R_0}\leq\frac{M_*}{H_I}\qquad\qquad\text{or}\qquad\qquad H_I\leq\frac{1}{\epsilon R_\perp}
\end{align}
to be compared with \eqref{TCCcon}. The order-one factor appearing in the four-dimensional Einstein frame can be further determined using
\begin{align}
    a\propto \hat{a}^{\frac{d+2}{2}}
\end{align}
so that, at the end of inflation, when the Weyl factor is equal to one, the four-dimensional and $(4+d)$-dimensional Hubble parameters are related by
\begin{align}
    h_e=\frac{d+2}{2}H_I.
\end{align}
Expressing the TCC in the four-dimensional Einstein frame then leads to
\begin{align}\label{eq:TCChigherdimEinstein}
    \frac{\hat{a}_e}{\hat{a}_0}\leq\frac{2}{d+2}\frac{M_*}{H_I}\qquad\qquad\text{or}\qquad\qquad H_I\leq\frac{2}{d+2}\frac{1}{\epsilon R_\perp}
\end{align}
which reproduces, as expected, \eqref{eq:TCCfactor} for $d=1$.

\subsection{The TCC in 6D inflation}

Let us now specialize the analysis to the case $d=2$. To this end, we consider the minimal setup of 6D inflation followed by a linear phase, $w=-1/3$, extending from the stabilization of the extra dimensions to late-time reheating. This choice maximizes the number of e-folds $N_4$ produced during this phase and therefore corresponds, once again, to the most favorable scenario.

The analysis proceeds as before: the scale invariance and horizon problem conditions must still be satisfied and are given by
\begin{align}
    2\pi^2 R_\perp e^{N_4+N_r}\geq\lambda_t\qquad\text{and}\qquad  D_{\text{LSS}}\leq e^{\frac{1}{2}N+N_4+N_r}\frac{1}{2H_I}
\end{align}
respectively, where we used $h_0=3H_IR_\perp/R_0$. Using the following expressions for the number of e-folds during inflation in the Einstein frame $N$ and during the subsequent 4D phase $N_4$:
\begin{align}
    e^{N}&=\left(\frac{R_\perp}{R_0}\right)^{2}=R_\perp M_p\frac{\epsilon^{2}}{\pi},\\
    e^{N_4}&=\left(\frac{h_r}{h_e}\right)^{-1}=\left(\sqrt{\frac{\pi^2g_*}{90}}\frac{T_r^2}{M_p}\frac{1}{2H_I}\right)^{-1},\label{eq:N46D}
\end{align}
where, in the expression for $N_4$, we used $h_e=2H_I$, while $N_r$ is still given by \eqref{eq:N_r}. The scale invariance condition then becomes
\begin{align}\label{eq:SI6D}
    \frac{H_I}{\text{GeV}}\geq2.6\times10^{-4}\,g_*^{1/2}\left(\frac{T_r}{\text{MeV}}\right)\left(\frac{R_\perp}{\mu\text{m}}\right)^{-1}
\end{align}
whereas the horizon problem condition reads
\begin{align}\label{eq:hor6D}
    \epsilon\geq4.3\times10^{-7}\,g_*^{1/2}\left(\frac{T_r}{\text{MeV}}\right)\left(\frac{R_\perp}{\mu\text{m}}\right)^{-1/2}.
\end{align}
They can be further combined to obtain
\begin{align}\label{eq:6DSI+hor}
    \epsilon\left(\frac{H_I}{\text{GeV}}\right)\geq1.1\times10^{-10}\,g_*\left(\frac{T_r}{\text{MeV}}\right)^{2}\left(\frac{R_\perp}{\mu\text{m}}\right)^{-3/2}.
\end{align}

We now turn back to the TCC. In the 4D Einstein frame, the TCC \eqref{eq:TCChigherdimEinstein} reads 
\begin{align}\label{eq:TCC6D}
    \frac{H_I}{\text{GeV}}\leq1.0\times10^{-10}\,\epsilon^{-1}\left(\frac{R_\perp}{\mu\text{m}}\right)^{-1}.
\end{align}
Compatibility of the TCC with \eqref{eq:6DSI+hor} therefore requires 
\begin{align}
    \left(\frac{R_\perp}{\mu\text{m}}\right)^{1/2}\geq1.1\,g_*\left(\frac{T_r}{\text{MeV}}\right)^{2}.
\end{align}

One should recall that collider constraints impose a lower bound on the six-dimensional gravity scale, $M_* \gtrsim 10~\text{TeV}$~\cite{Anchordoqui:2025nmb}. This translates into an upper bound on the compactification radius, namely $R_\perp \lesssim 1~\mu\text{m}$. We therefore adopt the benchmark value $R_\perp=1\,\mu\text{m}$, which yields the most conservative upper bound on $T_r$. In addition, the reheating temperature $T_r$ must remain below the normalcy temperature, which is given by
\begin{align}
    T_N\simeq4\left(\frac{R_\perp}{\mu\text{m}}\right)^{-1/2}\,\text{MeV}.
\end{align}
A detailed derivation can be found in~\cite{Anchordoqui:2026los}. For the benchmark value $R_\perp=1\,\mu\text{m}$, this implies $T_r\lesssim4\,\text{MeV}$, corresponding to $g_*\sim10$~\cite{Husdal:2016haj}. Altogether, we therefore end up with the constraint $T_r\lesssim0.3\,\text{MeV}$, which is in strong tension with the successful predictions of BBN.\footnote{One can check that including an additional phase of 6D expansion after inflation, in analogy with Sec.~\ref{sec:5Dalpha}, would further tighten this constraint by introducing an additional suppression factor $\alpha\leq1$ in the upper bound.} We therefore conclude that, at least within the minimal cosmological scenario considered here, 6D inflation is not phenomenologically viable once the TCC is imposed.

However, combining \eqref{eq:SI6D} and \eqref{eq:TCC6D}, we obtain
$\epsilon\lesssim10^{-8}$ for $T_r\sim 4\,\text{MeV}$, similarly to the case of $d=1$. We can therefore
consider the possibility that such a small initial value is generated through a
pre-inflationary phase, as discussed in the previous section. For a 6D
linear pre-inflationary phase, \eqref{eq:hor6D} is modified into
\begin{align}
    \epsilon\big(1-\ln(\epsilon)\big)\geq4.3\times10^{-7}\,g_*^{1/2}\left(\frac{T_r}{\text{MeV}}\right)\left(\frac{R_\perp}{\mu\text{m}}\right)^{-1/2}.
\end{align}
The 6D scenario can then become marginally viable for
$T_r\sim1\,\text{MeV}$, $H_I\sim1\,\text{MeV}$ and
$\epsilon\sim10^{-7}$.

\section{Comments of M-theory realisation}\label{sec:Mtheory}

An attractive realisation of the $d=1$ (dark dimension) setup was
proposed in~\cite{Schwarz:2024tet} within the 11D M-theory framework
describing the strong coupling limit of the Heterotic
string~\cite{Horava:1995qa,Witten:1996mz,Banks:1996ss}. The 11th dimension is compactified on an
interval of length $\pi R$ which vanishes in the weak string coupling
limit, with two 9-branes localised at its boundaries, each one
carrying an $E_8$ gauge group with a 10D gauge coupling fixed by the
11D gravity scale $M_{11}=l_{11}^{-1}$ up to a calculable
constant. The six extra dimensions are compactified on an internal
space of size $\rho$ defined so that its volume is $\rho^6$. Thus, one
dimension is special with respect to the others, along which the
observable world (embedded in one of the two $E_8$'s) is localised.\footnote{Concerns about proton stability~\cite{Reig:2025gch} have been addressed in~\cite{Blumenhagen:2026rgu}.} 

When the theory in the first $E_8$ sector is perturbative, the second
$E_8$ sector becomes non-perturbative at large dark dimension radii
($R_\perp$). To prevent its coupling 
from turning negative, this transition imposes a strict upper limit on the
size of the 11th dimension that 
rules out a radius of $R_\perp \sim 1\,\mu\text{m}$. This behavior is typical in orientifold compactifications, where
localised interaction couplings receive corrections that scale linearly with the size of a
large transverse extra dimension~\cite{Antoniadis:1998ax}. Preventing these divergences
generally requires local tadpole cancellation between branes and
orientifolds. Thus, certain non-geometric compactifications bypass this restriction entirely; in these specific cases, the correction to the second $E_8$ coupling vanishes, leaving the dark dimension radius ($R_\perp$) unconstrained~\cite{Caceres:1996is}.

Consequently, in this specific set-up there is a natural separation of
scales $l_{11}<l_*<\rho<R$, in terms of which our 5D effective theory
with gravity scale $M_*=l_*^{-1}$ and a  compact mesoscopic dimension of radius
$R_\perp$ is defined by the identification of scales:
\begin{align}
\label{Mtheory}
\alpha_G\sim \left(\frac{l_{11}}{\rho}\right)^6<1\quad;\quad  M_*^3=\rho^6 M_{11}^9
\quad;\quad M_p^2=\pi R_\perp\rho^6M_{11}^9\,,
\end{align}
where $\alpha_G$ is the 4D gauge coupling~\cite{Anchordoqui:2024ajk}. The initial condition $R=R_0$ at the beginning of inflation defines an initial 4D Planck scale $M_{p,0}=(R_0/R_\perp)^{1/2}M_p\sim\epsilon^{-1/2} M_*$, where $\rho$ is stabilised near $l_*$. 

Above $M_*$ and below $M_{p,0}$, both $\rho$ and $R$ are moduli and may evolve in time. Indeed, vacuum Einstein equations have Kasner solutions~\cite{Kasner:1940kz} that keep the total internal volume $(\pi R\rho^6)$, and thus $M_{p,0}$, fixed:
\begin{align}
\label{Kasner}
R(t)=R_0(t/t_0)^q\,,\quad \rho(t)=\rho_0(t/t_0)^p\quad;\quad q+6p=1\,,\quad q^2+6p^2=1\,,
\end{align}
implying $p=\pm{1/\sqrt{42}}$ and $q=\mp\sqrt{6/7}$. Thus, when $R$ expands the  other six radii shrink but in a much slower rate and vice versa. At $t=t_0$, $\rho_0^6M_{11}^9=M_*^3$, while in the past we consider the case where the 11th dimension was smaller and the other six larger according to \eqref{Kasner}, so that initially at $t=t_i$ one has:
\begin{align}
\label{KasnerM11}
t=t_i:\quad R_0/R_i = (\rho_i/\rho_0)^6 \quad\to\quad t=t_0:\quad M_*=\rho_0^2M_{11}^3\,.
\end{align}
It follows that at $t_i$ all radii could be at the same order within an order of magnitude larger than the fundamental scale $l_{11}$, explaining the smallness of $\epsilon$ at a later time $t_0$ where the size of the internal manifold $\rho_0$ is fixed at a high scale by some stabilisation mechanism.

\subsection{Linear expanding Universe}

\label{subsec:leU}

The identification of the dark dimension with the 11-th dimension of M-theory, which plays the role of the string coupling given by the dilaton, $g_s=e^\Phi\sim R_\perp$, motivates a possible role of the so-called linear dilaton solution during some period in the universe evolution~\cite{Myers:1987fv, Antoniadis:1988aa, Antoniadis:1988vi}. It consists in an exact string solution where the string frame metric is flat and the dilaton is linear in the time coordinate $X^0$, 
\begin{align}
\label{LD}
\Phi=QX^0\,;\quad g_s=e^\Phi\sim R_\perp/\sqrt{\alpha'}\,.
\end{align}
From the world-sheet point of view, it corresponds to a conformal field theory (CFT) where the time coordinate has a background charge $Q$ that lowers its central charge to $1-12Q^2$, thus increasing the critical dimension to $D>26$ (or $D>10$) of the bosonic (or fermionic) string (supercritical string) with $D-26=12Q^2$ (or $D-10=8Q^2$). The background \eqref{LD} is a solution of the string effective action in $D$ dimension:
\begin{align}
\label{stringaction}
S_\text{string}=\frac{1}{2}\int [d^DX]e^{-2\Phi}\left\{{\cal R}^{(D)}+4(\nabla\Phi)^2-4Q^2\right\}\,,
\end{align}
where ${\cal R}^{(D)}$ is the $D$-dimensional curvature scalar and we have set the Regge slope $\alpha'=1$.

After rescaling the metric from the string to the Einstein frame, one obtains a linearly expanding universe with the dilaton depending logarithmically in time:
\begin{align}
\label{LU}
ds^2=-dt^2+\left(\frac{2Qt}{D-2}\right)^2d\vec{x}^2\,;\quad \Phi=\frac{D-2}{2}\ln\frac{2Qt}{D-2}\,.
\end{align}
The exponent of the dilaton potential in the Einstein frame, after canonically normalising its kinetic term, becomes $2/\sqrt{D-2}$ which is the limiting lower bound~\cite{Etheredge:2022opl} according the asymptotic Swampland conjecture~\cite{Agmon:2022thq}. 
On the other hand, it is interesting to note that the string frame for
$D=10$ corresponds to the Einstein frame of M-theory containing one
extra dimension that plays the role of the dilaton, with the string
tree-level dilaton potential in \eqref{stringaction} replaced by
$H_3^2$ where $H_3$ is the field strength of the Neuveu-Schwarz (NS)
antisymmetric tensor descending from the M-theory 3-form with one
component along the 11-th dimension. Thus, upon compactification in 4
dimensions on a Calabi-Yau manifold a constant NS 3-form flux can
generate the dilaton potential leading to the linear dilaton
solution. Using the identification of M and string theory parameters:
\begin{align}
\label{M-string}
g_s=e^\Phi=(M_{11}R_\perp)^{3/2}\,;\quad M_s^2=M_{11}^3R_\perp\,,
\end{align}
where $M_s$ is the string scale and $M_{11}\sim 1/\rho$ from \eqref{Mtheory}, one can uplift the linear dilaton solution \eqref{LD} in the M-theory metric:
\begin{align}
\label{MLD}
ds^2_\text{M-theory}=e^{-2Q\tau/3}(-d\tau^2+d\vec{x}^2)+e^{4Q\tau/3}dy^2\,,
\end{align}
where $y$ denotes the 11-th dimension, while there is also a 4-form flux $G_{11ijk}$ with $i,j,k$ holomorphic indices on the Calabi-Yau manifold, generating the dilaton potential. Thus, when the size of the 11-th dimension expands, the universe contacts with half-rate of the expansion and vice-versa.

\section{Conclusions}
\label{sec:conclusion}

In this work, we investigated the implications of the TCC for higher-dimensional inflation. In this class of models, inflation simultaneously enlarges the compact extra dimensions from an initial size $R_0$ to their present value while expanding the three non-compact spatial dimensions sufficiently to solve the horizon problem. Higher-dimensional inflation therefore provides a unified framework relating two striking hierarchies in nature: the weakness of gravity and the large size of the observable Universe.

It was previously shown that this scenario naturally predicts an approximately scale-invariant spectrum of primordial scalar perturbations, recovered upon summation over the KK modes, provided the size of the extra dimensions is of order the micron scale. Combined with experimental bounds from short-range tests of Newton's inverse-square law, particle physics and astrophysical constraints, this leaves only two phenomenologically viable possibilities, corresponding to one or two large extra dimensions. The corresponding fundamental gravity scale is $M_*\sim10^9\,\text{GeV}$ for one extra dimension and $M_*\sim10\,\text{TeV}$ for two. These values naturally connect to the recent dark dimension proposal related to the smallness of the dark energy in the case $d=1$, and to the original large extra-dimension framework addressing the mass hierarchy problem in the case $d=2$.

In a companion paper~\cite{Anchordoqui:2026los}, we proposed a cosmological history connecting higher-dimen\-sional inflation to the onset of standard cosmology. It consists in an intermediate non-accelerating phase characterized by an effective equation-of-state parameter $w\ge -1/3$, followed by a phase of inflaton oscillations around the minimum of its potential, corresponding to $w_{os}=0$ for an approximately quadratic potential. Throughout this work, we focused on the most favorable case, $w=-1/3$, which maximizes the expansion during the intermediate phase.

The parameter space of the model is constrained by the simultaneous requirements of reproducing a nearly scale-invariant power spectrum of primordial fluctuations and of solving the horizon problem. Here, we investigated how these constraints are modified once the TCC is imposed. In higher-dimensional inflation, we showed that TCC takes the simple form
\begin{align}
H_I\lesssim\frac{1}{\epsilon R_\perp}\quad\text{with}\quad\epsilon=\frac{1}{R_0M_*}\,,
\end{align}
up to an order-one factor of $2/(d+2)$ when expressed in the 4D Einstein frame. This in turn constrains the inflaton mass $m$ to values incompatible with a
quadratic oscillation phase providing the subsequent cosmological evolution,
as it requires $m\lesssim10^{-9}\,\text{eV}$.

We therefore led to consider instead a cosmological history without the oscillatory
phase and derived the minimal requirements for satisfying TCC. We then found that
TCC imposes that both the initial radius $R_0$ and the Hubble radius of higher-dimensional
inflation $H_I^{-1}$ must be significantly larger than the
fundamental length scale $M_*^{-1}$:
\begin{align}
    \epsilon=\frac{1}{R_0 M_*}\lesssim10^{-8}
    \qquad\qquad\text{and}\qquad\qquad
    \frac{H_I}{M_*}\lesssim 10^{-9}
\end{align}
for $d=1$. For $d=2$, TCC strongly constrains the scenario, requiring a reheating temperature $T_r\lesssim0.3\,\text{MeV}$, which is in strong tension with the successful predictions of BBN and therefore challenges its phenomenological viability. The smallness of the Hubble constant of inflation implies also a tuning of the inflaton potential since it comes with a strong suppression of primordial gravitational waves
\begin{align}
    r\lesssim10^{-19}\,,
\end{align}
although better than the 
\begin{align}
    r\lesssim10^{-50}
\end{align}
imposed by the Higuchi bound in case of 4D inflation in the presence of a stabilized large extra dimension. However, the tensor-to-scalar ratio is given by $r=16\,\varepsilon_{V}^{(4)}$ in 4D and by $r=24\,\varepsilon_{V}^{(5)}$ in 5D with
\begin{align}
    \varepsilon_{V}^{(4)}=\frac{M_p^2}{2}\left(\frac{V'}{V}\right)^2\qquad\qquad\text{and}\qquad\qquad \varepsilon_{V}^{(5)}=\frac{3M_*^3}{4}\left(\frac{V'}{V}\right)^2\,,
\end{align}
implying a 5D $V'/V\lesssim 10^{-10}$ in Planck units, where prime
denotes derivative with respect to $\phi$. Note that this is much
better than the $V'/V\lesssim 10^{-25}$ of the 4D set up.

We finally discussed the possibility that such small values of $R_0^{-1}$ and $H_I$ compared to $M_*$
could arise dynamically through a pre-inflationary expansion of the Universe. This could be realised in the context of heterotic M-theory that admits Kasner type solutions where the 11-th dimension transverse to the branes at the boundaries expands while the remaining six along the branes shrink at a rate much slower than the expansion of the former, so that the desired hierarchy can arise naturally from an initial condition where the size of all dimensions is comparable to an order of magnitude higher that the fundamental length. On the other hand, the identification of the dark dimension with the 11-th dimension of M-theory, which plays the role of the string coupling given by the dilaton, motivates a possible role of the linear dilaton background which is an exact string solution and describes in the Einstein frame a linear expansion of the universe in any dimension and at any period of its evolution, before or after higher-dimensional inflation, thus motivating the various possibilities we discussed in this paper. An additional important consequence of a pre-inflationary period of linear expansion is the weakening of the horizon constraint leading to a higher upper bound for the inflationary Hubble parameter by 2-3 orders of magnitude and consequently of $r$ by 6-9 orders of magnitude to $r\lesssim 10^{-10}$. This also implies much less required tuning of the inflaton potential $V'/V\lesssim 10^{-5}$ in 5D.

\section*{Acknowledgements}

We are grateful to Alek Bedroya for valuable discussions. I.A. and
L.A.A. would like to thank the hospitality of the Center for Cosmology
and Particle Physics in NYU where part of this work was performed. The
work of L.A.A. is supported by the U.S. National Science Foundation
(NSF Grant PHY-2412679), he extends his appreciation to the Harvard
Swampland Initiative for their warm hospitality and for providing a
stimulating environment for productive discussions. The research of
I.A. was supported in part by the Higher Education and Science
Committee of MESCS RA (Research Project N 24RL-1C036).

\appendix

\bibliography{biblio}

\end{document}